\documentclass[aps,prl,nopacs,twocolumn,preprintnumbers,notitlepage,amsmath,amssymb,superscriptaddress]{revtex4-2}
\usepackage[colorlinks=true,allcolors=blue]{hyperref} 
\usepackage{amssymb}
\usepackage{amsmath}
\usepackage{color}
\usepackage[pdftex]{graphicx}
\usepackage{bm}  
\usepackage{graphicx}
\usepackage{amscd}
\usepackage{epsfig}
\usepackage{epstopdf}
\usepackage{enumerate}
\usepackage{mathtools}
\usepackage{notes2bib}

\newcommand{\ds}{\displaystyle}
\newcommand{\beq}{\begin{equation}}
\newcommand{\eeq}{\end{equation}}
\newcommand{\beqa}{\begin{eqnarray}}
\newcommand{\eeqa}{\end{eqnarray}}
\newcommand{\bem}{\begin{math}}
\newcommand{\eem}{\end{math}}

\newcommand{\pa}[1]{\partial_{#1}}

\newcommand{\cR}{{\mathcal R}}
\newcommand{\cC}{{\mathcal C}}
\newcommand{\kT}{{k_{\rm B} T}}
\newcommand{\cP}{{\mathcal P}}
\newcommand{\cG}{{\mathcal G}}
\newcommand{\cS}{{\mathcal S}}

\newcommand{\bcM}{{\bm{\mathcal M}}}
\newcommand{\bcZ}{{\bm{\mathcal Z}}}
\newcommand{\bcY}{{\bm{\mathcal Y}}}
\newcommand{\cM}{{\mathcal M}}
\newcommand{\cZ}{{\mathcal Z}}
\newcommand{\cY}{{\mathcal Y}}

\newcommand{\bfr}{{\bm r}}

\newcommand{\bfh}{{\bm h}}

\newcommand{\bfq}{{\bm q}}

\newcommand{\bfx}{{\bm x}}
\newcommand{\bff}{{\bm f}}
\newcommand{\bfv}{{\bm v}}

\newcommand{\bfR}{{\bm R}}

\newcommand{\bnabla}{{\bm \nabla}}

\newcommand{\bxi}{{\bm \xi}}
\newcommand{\bphi}{{\bm \phi}}

\newcommand{\aver}[1]{\left\langle {#1}\right\rangle}

\begin{document}

\title{Hydrodynamically consistent many-body Harada-Sasa relation}

\author{Ramin Golestanian}
\email{ramin.golestanian@ds.mpg.de}
\affiliation{Max Planck Institute for Dynamics and Self-Organization (MPI-DS), 37077 G\"ottingen, Germany}
\affiliation{Rudolf Peierls Centre for Theoretical Physics, University of Oxford, Oxford OX1 3PU, United Kingdom}

\date{\today}

\begin{abstract}
The effect of hydrodynamic interactions on the non-equilibrium stochastic dynamics of particles -- arising from the conservation of momentum in the fluid medium -- is examined in the context of the relationship between fluctuations, response functions, and the entropy production rate. The multiplicative nature of the hydrodynamic interactions is shown to introduce subtleties that preclude a straightforward extension of the Harada-Sasa relation. A generalization of the definitions involved in the framework is used to propose a new form of the relation applicable to systems with hydrodynamic interactions. The resulting framework will enable characterization of the non-equilibrium properties of living and active matter systems, which are predominantly in suspensions.
\end{abstract}

\maketitle

Since the inception of the Onsager regression hypothesis \cite{onsager2}, understanding the violation of the formal correspondence between fluctuations and response functions as observed in a variety of non-equilibrium systems \cite{LetPeliti1997,LetDean1997,Golestanian2002,Crisanti2003,haradaEqualityConnectingEnergy2005,haradaEnergyDissipationViolation2006,Speck2006,Maes2009,Parrondo2009,Harada2009,Turlier2016,Abah2016,Maes2020} has guided many conceptual developments in non-equilibrium statistical physics. A particularly elegant breakthrough has been provided by the establishment of a formal relationship between the spectral summation of the difference between correlation and response, and the rate of energy dissipation, which quantifies how far away from equilibrium a given system operates \cite{haradaEqualityConnectingEnergy2005,haradaEnergyDissipationViolation2006,Harada2009}.

With the advent of active matter \cite{Gompper2020}, new horizons have opened in theoretical and experimental studies of non-equilibrium systems where questions pertaining to such relationships, as well as the degree of non-equilibrium activity and entropy production, can be systematically investigated \cite{battle2016broken,nardini2017entropy,pietzonka2017entropy,shankar2018hidden}. These biological and synthetic active matter systems typically produce mechanical activity in the form of propulsion or delivering work as fuelled by a locally available source of free energy \cite{julicher1997modeling,ramaswamy2010,mugnai2020theoretical,borsley2022chemical,Pumm2022,Shi2022,Shi2023,Golestanian2019phoretic}. A key question in quantifying the non-equilibrium character of such active matter systems corresponds to the amount of hydrodynamic dissipation, which is inherently related to the faithful implementation of momentum conservation, from the nano-scale \cite{RG2008,Mike3SS2024} to the micro-scale \cite{Babak2021,daddi2023minimum,speck}.

The Harada-Sasa relation connects the rate of entropy production ${\dot \sigma}$ in a driven non-equilibrium system under stationary-state conditions to the spectral sum of the violation of the fluctuation-dissipation theorem, i.e. the difference between the velocity correlation function $\cC(t)$ and the response function $\cR(t)$ in the frequency domain. More specifically, for a single particle (in 1D) with friction coefficient $\zeta$ immersed in a medium with temperature $T$, the Harada-Sasa relation is written as \cite{haradaEqualityConnectingEnergy2005,haradaEnergyDissipationViolation2006,Harada2009}
\beq
T {\dot \sigma}=\zeta \bar{V}^2+ \zeta\int_{-\infty}^{\infty} \frac{d \omega}{2 \pi} \left[\tilde{\cC}(\omega)-2 \kT \tilde{\cR}'(\omega)\right],\label{eq:HS-def-1}
\eeq 
where $\bar{V}$ is the mean velocity, $k_{\rm B}$ is the Boltzmann constant, and $\tilde{\cR}'(\omega)=\frac{1}{2}[\tilde{\cR}(\omega)+\tilde{\cR}(-\omega)]$ is the real part of the response function. An intuitive way to think about Eq. (\ref{eq:HS-def-1}) is to interpret it as an energy flux balance equation in which the work done by the viscous drag force that is converted into heat and released to the medium minus the work done by the stochastic Brownian force that is extracted from the thermal bath results in a net flux that balances the work done by the external driving force and released into the medium. As the energy of the system will not change in stationary state, this is effectively a reflection of the first law of thermodynamics, which governs a corresponding extension of the Onsager regression hypothesis. A central premise in the derivation of this relation is the assumption that the dissipation occurs via a local mechanism, which is not consistent with hydrodynamic dissipation in a momentum-conserving background fluid, as evidenced e.g. by the emergence of the hydrodynamic long-time tails \cite{Alder1967,Zwanzig1970} instead of the exponential crossover between the inertial and viscous regimes postulated by Langevin in his description of the Brownian motion \cite{Langevin_1908}.

In this Letter, we set out to develop a hydrodynamically consistent extension of the Harada-Sasa relation, which can be applied to non-equilibrium suspensions in steady-state conditions. We achieve this by enforcing momentum conservation both on the particles in the suspension and on the medium itself, which is typically an incompressible viscous fluid that satisfies the Stokes equation. After eliminating the dynamics of the fluid medium, the resulting many-body stochastic dynamics constitutes a process with multiplicative noise, and should therefore be handled with care. Our approach is rooted in the correct physical characterization of the fluctuations of the non-equilibrium medium, and as such proves to be free of foundational problems associated with spurious drift terms that otherwise need to be augmented artificially. 

We observe that the structure of the many-body dynamics with the multiplicative noise is such that the term responsible for quantifying the breakdown of time-reversal symmetry -- that gives the entropy production rate -- is independent of the hydrodynamic interactions. This might be construed as suggesting that a straightforward extension of the Harada-Sasa relation for the many-body dynamics could hold independently of the hydrodynamic coupling between the particles. We uncover that this intuition turns out to be untrue, due to the subtle involvement of the hydrodynamic couplings in the relationship between the correlation functions and the response functions. We derive the correct form for the Harada-Sasa relation in the presence of hydrodynamic interactions, which involves appropriately generalized correlation functions and response functions.

We consider the many-body stochastic dynamics of $N$ colloidal particles in $d$ dimensions, with the $\alpha$th particle ($\alpha \in \{1 ,\ldots,N\}$) being described by position $\bfr^{\alpha}(t)$ and force $\bff^{\alpha}(t)$. Here, we consider the most general case in which the individual force acting on a given particle can depend on the position of the particle itself (arbitrary external force) as well as the positions of the other particles (arbitrary interactions, including many-body interactions). The generality of the choice for the forces will enable us to apply this formalism to a variety of active and driven systems (see the discussion below for more details). We can quantify the entropy production rate of the system as the average of the stochastic energy dissipation rate as follows
\beq
T {\dot \sigma}=\aver{\dot{r}^\alpha_i(t) f^{\alpha}_i \left(\{\bfr^{\nu}(t)\}\right)},\label{eq:entropy-def-1} 
\eeq
where summation over repeated indices is assumed ($i \in \{1 ,\ldots,d\}$). The averaging is to be performed using the corresponding path probability distribution weight, which reads
\beq
{\mathsf P}[\{\bfr^{\alpha}(t)\}]=\frac{1}{Z}\Big(\det \cM^{\alpha \beta}_{i j} \Big)^{-1/2} \exp\Big(-\beta\cS_{\rm OM}[\{\bfr^{\alpha}(t)\}]\Big),\label{eq:P-5}
\eeq
with the Onsager-Machlup action given as
\beq
\cS_{\rm OM}=\frac{1}{4} \int d t\; \cZ^{\alpha \beta}_{i j} \Big(\dot{r}^{\alpha}_i-\cM^{\alpha \gamma}_{i k}  f^{\gamma}_k\Big)  \Big(\dot{r}^{\beta}_j-\cM^{\beta \delta}_{j l}  f^{\delta}_l\Big).
\label{eq:S-OM-1}
\eeq
Here, $\bcZ(\{\bfr^{\alpha}(t)\})$ is the friction tensor, and $\bcM(\{\bfr^{\alpha}(t)\})$ is the mobility tensor, with the two being subject to the relation $\bcZ=\bcM^{-1}$. Moreover, $Z$ is a normalization constant, and $\beta \equiv 1/(\kT)$ [see Appendix A for the derivation of Eq. (\ref{eq:P-5})]. Note that our use of the Onsager-Machlup action is not based on a phenomenological choice but rather on a systematic derivation that eliminates the fluid degrees of freedom, which are governed by hydrodynamic fluctuations in the Stokes regime.

An inspection of Eq. (\ref{eq:S-OM-1}) when re-written as follows
\beq
\cS_{\rm OM}=\frac{1}{4} \int d t \left[\cZ^{\alpha \beta}_{i j} \dot{r}^{\alpha}_i \dot{r}^{\beta}_j -2 \,\dot{r}^\alpha_i f^{\alpha}_i +\cM^{\alpha \beta}_{i j} f^{\alpha}_i f^{\beta}_j\right],
\label{eq:S-OM-2}
\eeq
reveals that the term in the Onsager-Machlup action that quantifies the breakdown of time-reversal symmetry is independent of the mobility and friction tensors, and that it directly leads to the local definition of entropy production rate as given in Eq. (\ref{eq:entropy-def-1}). This is related to the fact that the Fokker-Planck equation for the many-particle probability $\cP(\{\bfx^\alpha\},t)$ associated with the above Onsager-Machlup action, namely ($\pa{i}^{\alpha}\equiv \partial/\partial x^\alpha_i$)
\beq
\pa{t} \cP(\{\bfx^\alpha\},t) + \pa{i}^{\alpha} \left[\cM^{\alpha \beta}_{i j} \Big(f^{\beta}_{j} \, \cP-\kT \pa{j}^{\beta} \cP\Big)\right]=0,\label{eq:FP-hydro}
\eeq
guarantees equilibration without any involvement of the mobility tensor, when $f^{\beta}_{j}=-\pa{j}^{\beta}U$ for some scalar potential energy $U(\{\bfx^\alpha\})$. In this case, the many-body multiplicative coupling of the mobility tensor only affects the transient dynamics through the relaxation to equilibrium as described by $\cP_{\rm eq}(\{\bfx^\alpha\}) \propto \exp\left(-\beta U\right)$. In light of these observations, one naturally wonders if the same fate applies to the Harada-Sasa relation, in the sense that the local definition of entropy production rate given in Eq. (\ref{eq:entropy-def-1}) still quantifies the spectral sum of the degree of violation of the fluctuation-dissipation relation, as naturally defined in terms of the difference between the correlation function and the response function of the many-body system in the frequency domain. This is what we will now investigate. 

To develop the framework for the derivation of the Harada-Sasa relation, we start with the average value of the velocity of a given particle 
\beq
\aver{\dot{r}^\alpha_i(t)}=\int \prod_\alpha {\mathcal D} \bfr^\alpha(\tau) {\mathsf P}[\{\bfr^{\alpha}(\tau)\}]\dot{r}^\alpha_i(t). \label{eq:rdot-def-1} 
\eeq
We next consider a set of external forces $\bff^{\beta}_{\rm ext}$ that act on all the particles, and use them to define the response functions
\beq
\cR^{\alpha \beta}_{i j} (t,t')=\cR^{\alpha \beta}_{i j} (t-t') \equiv \left.\frac{\delta \aver{\dot{r}^\alpha_i(t)}_{\bff^{\beta}_{\rm ext}}}{\delta f^{\beta}_{j,{\rm ext}}(t')}\right|_{\bff^{\beta}_{\rm ext}={\bf 0}}. \label{eq:R-def-1}
\eeq
The first equality holds because we consider the system to be in steady-state conditions, and as such, expect time-translation invariance to hold. The calculation yields
\beqa
&&\cR^{\alpha \beta}_{i j} (t-t')= \frac{\beta}{2} \left[\bar{V}^\alpha_i \bar{V}^\beta_j+\cC^{\alpha \beta}_{i j} (t-t')\right]\nonumber \\
&&\hskip0.7cm- \frac{\beta}{2} \aver{\dot{r}^\alpha_i(t) \cM^{\beta \gamma}_{j k}\left(\{\bfr^{\nu}(t')\}\right)  f^{\gamma}_k \left(\{\bfr^{\nu}(t')\}\right)}, \label{eq:R-res-1}
\eeqa
where we have defined the average velocities as $\bar{V}^\alpha_i \equiv \aver{\dot{r}^\alpha_i(t)}$, and the correlation functions as 
\beq
\cC^{\alpha \beta}_{i j} (t-t') \equiv \aver{\big[\dot{r}^\alpha_i(t)-\bar{V}^\alpha_i\big]\big[\dot{r}^\beta_j(t')-\bar{V}^\beta_j\big]}. \label{eq:C-def-1}
\eeq
It is straightforward to show that at equilibrium, where the forces are conservative ($f^{\beta}_{j}=-\pa{j}^{\beta}U$), the average fluxes vanish ($\bar{V}^\alpha_i=0$), and the system has both time-reversal and time-translation symmetries, Eq. (\ref{eq:R-res-1}) leads to the celebrated fluctuation-response relation
\beq
\kT \cR^{\alpha \beta}_{{\rm eq},i j} (t-t')=\Theta(t-t') \, \cC^{\alpha \beta}_{{\rm eq},i j} (t-t'),\label{eq:FDT-time}
\eeq
where $\Theta(t)$ is the Heaviside step function, which enforces causality. In the frequency domain, Eq. (\ref{eq:FDT-time}) can be expressed as follows
\beq
\tilde{\cC}^{\alpha \beta}_{{\rm eq},i j}(\omega)=2 \kT \tilde{\cR}'^{\alpha \beta}_{{\rm eq},i j}(\omega),\label{eq:FDT-freq}
\eeq
where $\tilde{\cR}'^{\alpha \beta}_{{\rm eq},i j}(\omega)$ corresponds to the real part of the response function. 

Back to the general many-body non-equilibrium steady-state described by Eq. (\ref{eq:R-res-1}), we proceed by constructing the symmetrized form of the response functions $\cR^{\alpha \beta}_{i j} (t-t')+\cR^{\beta \alpha}_{j i} (t'-t)$ and taking the trace of the tensor. In the next step, we aim to take the limit $t \to t'$, which is a subtle calculation that needs to be performed using an appropriate discretization of the time steps \cite{haradaEqualityConnectingEnergy2005,haradaEnergyDissipationViolation2006}. Following this procedure and using a shorthand $\cR^{\alpha \alpha}_{i i} (0)\equiv \frac12 \lim_{\Delta t \to 0^+} \left[\cR^{\alpha \alpha}_{i i} (\Delta t)+\cR^{\alpha \alpha}_{i i} (-\Delta t)\right]$ (also for the correlation function), we arrive at the following expression
\beq
2 \cR^{\alpha \alpha}_{i i} (0)= \beta \left[\bar{V}^\alpha_i \bar{V}^\alpha_i+\cC^{\alpha \alpha}_{i i} (0)\right]-\beta {\cal J},\label{eq:RC0-1}
\eeq
where 
\beq
{\cal J}\equiv\aver{\dot{r}^\alpha_i(t) \cM^{\alpha \gamma}_{i k}\left(\{\bfr^{\nu}(t)\}\right)  f^{\gamma}_k \left(\{\bfr^{\nu}(t)\}\right)},\label{eq:J-def-1}
\eeq
plays the role of a generalized flux. We can re-arrange Eq. (\ref{eq:RC0-1}) and express it as follows
\beq
{\cal J}=\bar{V}^\alpha_i \bar{V}^\alpha_i+\int_{-\infty}^{\infty} \frac{d \omega}{2 \pi} \left[\tilde{\cC}^{\alpha \alpha}_{i i} (\omega)-2 \kT \tilde{\cR}'^{\alpha \alpha}_{i i} (\omega)\right].\label{eq:RComega-1}
\eeq 
Since ${\cal J}$ is not directly related (or proportional) to the entropy production rate as defined in Eq. (\ref{eq:entropy-def-1}) (as $\aver{{\dot r} \cM f} \neq \aver{\cM} \aver{{\dot r} f}$ in the general case), we observe that the Harada-Sasa relation does not hold in its original form in such a many-body system.

It is, however, possible to construct a generalization of the Harada-Sasa relation for the many-body dissipative dynamics, by using appropriately generalized definitions for the response function and the correlation function. To achieve this, we revisit the starting point of the calculation [Eq. (\ref{eq:rdot-def-1})] but instead start with $\aver{ \cZ^{\alpha \beta}_{i j}\left(\{\bfr^{\nu}(t)\}\right) \dot{r}^\beta_j(t)}$ and define the generalized response functions  
\beq
\cR^{\alpha \gamma}_{g, i k} (t,t') \equiv \left.\frac{\delta \aver{ \cZ^{\alpha \beta}_{i j}\left(\{\bfr^{\nu}(t)\}\right) \dot{r}^\beta_j(t)}_{\bff^{\gamma}_{\rm ext}}}{\delta f^{\gamma}_{k,{\rm ext}}(t')}\right|_{\bff^{\gamma}_{\rm ext}={\bf 0}}. \label{eq:Rg-def-1}
\eeq
Note that the response function is still translationally invariant in time in stationary state, and that explicit use of $(t,t')$ in the argument is meant to signify the time points used in the definition for the different quantities. Following the same steps of the calculation and defining the generalized correlation functions 
\beq
\cC^{\alpha \gamma}_{g,i k} (t,t') \equiv \aver{\cZ^{\alpha \beta}_{i j}\left(\{\bfr^{\nu}(t)\}\right)\dot{r}^\beta_j(t) \dot{r}^\gamma_k(t')},\label{eq:Cg-def-1}
\eeq
we obtain the following identity
\beqa
&&\cR^{\alpha \gamma}_{g,i k} (t,t')= \frac{\beta}{2} \cC^{\alpha \gamma}_{g,i k} (t,t')\nonumber \\
&&-\frac{\beta}{2} \aver{\cZ^{\alpha \beta}_{i j}(\{\bfr^{\nu}(t)\}) \dot{r}^\beta_j(t) \cM^{\gamma \delta}_{k l}(\{\bfr^{\nu}(t')\})  f^{\delta}_l(\{\bfr^{\nu}(t')\})}.\nonumber \\ \label{eq:Rg-res-1}
\eeqa
We now construct the symmetrized form of the generalized response functions $\cR^{\alpha \gamma}_{g,i k} (t,t')+\cR^{\gamma \alpha}_{g,k i} (t',t)$ and take the trace of the tensor. Next, we take the limit $t \to t'$ as described above, and use the shorthand as previously defined to write the resulting expression as
\beq
2 \cR^{\alpha \alpha}_{g,i i} (t,t)= \beta\cC^{\alpha \alpha}_{g,i i} (t,t)- \beta\aver{\dot{r}^\alpha_i(t) f^{\alpha}_i \left(\{\bfr^{\nu}(t)\}\right)},\label{eq:RgCg0-1}
\eeq
which can be re-written as follows
\beq
T {\dot \sigma}=\int_{-\infty}^{\infty} \frac{d \omega}{2 \pi} \left[\tilde{\cC}'^{\alpha \alpha}_{g,i i} (\omega)-2 \kT \tilde{\cR}'^{\alpha \alpha}_{g,i i} (\omega)\right].\label{eq:RgCg-omega-1}
\eeq
Note that in this case $\tilde{\cC}^{\alpha \alpha}_{g,i i} (\omega)$ can have both real and imaginary parts, because in general $\cC^{\alpha \alpha}_{g,i i} (t,t') \neq \cC^{\alpha \alpha}_{g,i i} (t',t)$.
The above result [Eq. (\ref{eq:RgCg-omega-1})] is our proposed generalization of the Harada-Sasa relation for a many-body system with hydrodynamic interactions. 

Remarkably, we can still think about Eq. (\ref{eq:RgCg-omega-1}) using the same intuition as originally described, namely, as a balance between the work done by the viscous drag forces, the contribution from the stochastic Brownian forces, and the work done by the active driving forces. The difference as compared to Eq. (\ref{eq:HS-def-1}) is that the drag force on the $\alpha$th particle depends on the velocities of all other particles, and this needs to be consistently taken into consideration as done in Eqs. (\ref{eq:Rg-def-1}) and (\ref{eq:Cg-def-1}). In this generalized formulation, $\cC^{\alpha \alpha}_{g,i i} (t,t')$ gives the correlation between the viscous drag forces and the velocities, and the fact that $\cC^{\alpha \alpha}_{g,i i} (t,t') \neq \cC^{\alpha \alpha}_{g,i i} (t',t)$ can be understood as a manifestation of the non-reciprocal nature of hydrodynamic interactions, which can lead to the emergence of exotic behaviour in active matter \cite{Ramin_10.1051epn2024305}. It is important to highlight that this intuitive picture is, strictly speaking, valid because we have chosen not to subtract out the average velocity in the generalized formulation [see the difference between Eqs. \eqref{eq:C-def-1} and \eqref{eq:Cg-def-1}] as it is done in the original formulation of the Harada-Sasa relation.

To demonstrate how the hydrodynamic interactions influence the calculations in practice, we consider a specific example of a two-sphere dumbbell, which is commonly used in the literature \cite{Dill1983,Frankel1989}, for example, to describe the dynamics of hydrodynamically interacting proteins and enzymes \cite{Illien2017,AdelekeLarodo2019,AgudoCanalejo2020}. We consider two identical spheres of radius $a$ described by positions $x^1$ and $x^2$, which can be converted to the separation $x=x^2-x^1$ and centre of friction $X=(x^1+x^2)/2$ coordinates, as they move (along the $x$-direction) under the influence of the forces $f^1=F-f(x)$ and $f^2=F+f(x)$; see Appendix B for more details. We find the following expression for the probability distribution
\beq
P[X(t),x(t)]=\frac{1}{Z} \prod_t \frac{1}{\sqrt{\mu(x(t))^2-\lambda(x(t))^2}} \, e^{-\beta S_{\rm OM}},\label{eq:P-x-X}
\eeq
where $S_{\rm OM}=\int_t \Big[{\zeta \dot{X}^2}/{(\mu+\lambda)}-2F \dot{X}+\frac{1}{\zeta}(\mu+\lambda) F^2+\frac{\zeta}{4} {\dot{x}^2}/{(\mu-\lambda)}-\dot{x} f(x(t))+\frac{1}{\zeta}(\mu-\lambda) f(x(t))^2\Big]$, and the components of the mobility tensor $\mu\left(x(t)\right)$ and $\lambda\left(x(t)\right)$ are functions of the separation between the two spheres. We can now calculate the quantities of interest according to the definitions given in Eqs. \eqref{eq:entropy-def-1}, \eqref{eq:Rg-def-1}, and \eqref{eq:Cg-def-1}, and express them in terms of correlations of $X(t)$ and $x(t)$. Since the $X(t)$ distribution is effectively Gaussian with time-dependent parameters, we can calculate the marginal averages and correlations and recast the expressions of interest in terms of averages of correlation function involving $x(t)$ alone; see Appendix B.

\begin{figure}[t]
\centering
\includegraphics[width=\columnwidth]{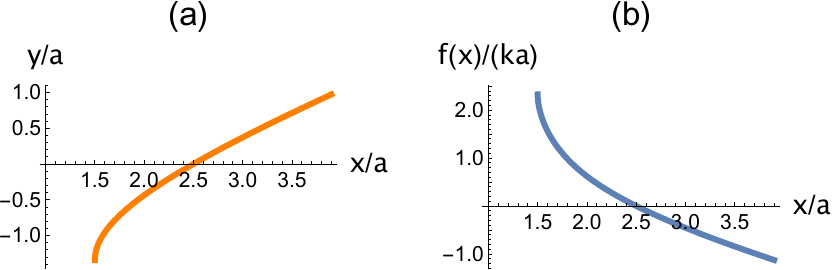}
\caption{Coordinate transformation (a) and the non-linear spring force-extension relation (b) for $\mu(x)=1$ and $\lambda(x)=3 a/(2 |x|)$, which correspond to the far-field form of the Oseen tensor (see Appendix A), plotted for $b=2.5 a$.}\label{fig:y-and-fx}
\end{figure}

To perform the calculations of the correlation functions, we make use of the coordinate transformation $y(x)=\frac{1}{2} \int_{b}^{x} d x_1/\sqrt{\mu(x_1)-\lambda(x_1)}$ and consider a non-linear spring acting between the two spheres with the force law $f(x)=-k y(x)/\sqrt{\mu(x)-\lambda(x)}$; see Fig. \ref{fig:y-and-fx} for plots of the coordinate transformation and the non-linear spring force law in the far-field approximation. In terms of the new coordinate, the system admits a Gaussian distribution of the form $P=\frac{1}{Z} \exp\left\{-\frac{\beta \zeta}{2} \int_{\omega} (\omega^2+k^2/\zeta^2) |y(\omega)|^2\right\}$, which enables perturbative calculation of the non-linear correlation functions, which appear in the generalized response function as follows
\beqa
\cR^{\alpha \alpha}_{g} (t,t')&=&\delta(t-t')+\beta \zeta \aver{\frac{g\left(y(t')\right)}{g\left(y(t)\right)} \dot{y}(t) \dot{y}(t')} \nonumber \\
&&+\beta k \aver{\frac{g\left(y(t')\right)}{g\left(y(t)\right)} \dot{y}(t) y(t')},
\eeqa
and, in the generalized correlation function as follows
\beqa
\cC^{\alpha \alpha}_{g} (t,t')&=&2 \kT\delta(t-t')+\frac{2  F^2}{\zeta} \aver{\mu+\lambda}\nonumber \\
&&+2 \zeta \aver{\frac{g\left(y(t')\right)}{g\left(y(t)\right)} \dot{y}(t) \dot{y}(t')},
\eeqa
where $g(y)\equiv \sqrt{\mu(x(y))-\lambda(x(y))}$. Using the Taylor expansion $g(y)=g_0+g_1 y+g_2 y^2 + \cdots$ and using Wicks theorem to evaluate the Gaussian correlations, we obtain
\beqa
&&\cR^{\alpha \alpha}_{g} (t,t')\simeq-\frac{k}{\zeta} e^{-k(t-t')/\zeta} \nonumber \\
&&\hskip1.3cm \times \Big[1+\frac{\kT}{2k}\left(\frac{g_1^2}{g_0^2}\right)\left(1-2  e^{-2 k(t-t')/\zeta}\right)\Big],
\eeqa
for $t>t'$, $\cR^{\alpha \alpha}_{g}(t,t')=0$ for $t<t'$, $T \dot{\sigma}=\frac{2  F^2}{\zeta} \aver{\mu+\lambda}$, and
\beqa
&&\cC^{\alpha \alpha}_{g} (t,t')\simeq 4 \kT\delta(t-t')+\frac{2  F^2}{\zeta} \aver{\mu+\lambda}\nonumber \\
&&\hskip.1cm -\kT \frac{k}{\zeta} e^{-k|t-t'|/\zeta}\Big[1+\frac{\kT}{2k}\left(\frac{g_1^2}{g_0^2}\right)\left(1-2  e^{-2 k|t-t'|/\zeta}\right)\Big],\nonumber \\
\eeqa
to the lowest non-trivial order. This example illustrates the subtleties involved in the calculations, due to the non-linearities introduced in the stochastic process.


Our generalization of the Harada-Sasa relation highlights the importance of a comprehensive mechanistic account of the modes of dissipation that govern the dynamics, as a guiding principle that can be helpful in understanding any non-equilibrium system. The above formulation can be readily applied to active systems, using appropriate implementation of active forces $\bff^{\alpha}(t)$ as the fundamental active components, while satisfying the appropriate force-free and torque-free constraints at the level of each active agent. Micro-swimmers can be described via oscillating force-multipoles \cite{Najafi2004} and coarse-grained into static force-dipoles beyond the time-scale associated with the swimming-gait cycle. Subsequently, this enables the appropriate pusher-puller classification \cite{RGAA2008}, which can be further coarse-grained to predict instabilities in micros-swimmer suspensions \cite{Simha2002,shelley2008,Baskaran2009,Tannie2010}. Active systems with the so-called slip boundary condition on the surface velocity -- such as phoretic micro-swimmers and squirmers -- can also be adapted within the current framework with some considerations. Phoretic micro-swimmers can be described at the microscopic level via an imbalance in the distribution of force-dipoles around the surfaces of the colloidal particles \cite{Golestanian2019phoretic}. We note that such a microscopic level of description is important when we aim to take a full account of the hydrodynamic dissipation in the system. The implementation of the surface-slip boundary condition for squirmer models necessitates the computation of the forces that can lead to such boundary conditions \cite{Brady1988,Pedley2008,Fielding2014}, which can be used to establish the connection to the framework presented here.

In conclusion, we have developed a framework that allows us to relate the rate of entropy production to the difference between appropriately defined observables that characterize correlations and response in many-body stochastic dynamics of particles that are immersed in a background fluid that is subject to the conservation of momentum. The framework can be extended in a variety of ways, including its application to field theories that incorporate a background momentum-conserving fluid, such as model H dynamics \cite{halperin1977} and its active generalization \cite{activeH2015}. Hydrodynamic consistency is expected to provide the appropriate platform for the characterization and quantification of the non-equilibrium properties of living and active matter systems, as well as driven non-equilibrium processes that have a significant involvement of the background fluid, such as sedimentation.

\textit{Appendix A: Many-body Onsager-Machlup action for particles with hydrodynamic interactions.---}We aim to setup a framework that takes account of the conservation of momentum for the individual particles and for the background fluid medium in which the particles reside. Our calculations concern the long-time behaviour of the system, where it is effectively governed by frictional dynamics rather than inertial dynamics, both for the particles and for the fluid medium. In this limit, which can be formally achieved by taking the vanishing limit for the particle masses and the fluid mass density, momentum conservation translates into instantaneous force balance for the particles and instantaneous stress balance at every point in the medium (i.e. the divergence of the stress tensor vanishes). This standard method can be shown to emerge from a more general calculation performed on the full dynamics that contains inertia, by taking the appropriate limit that gives the long-time behaviour.

We now derive the path probability distribution weight for the many-body stochastic dynamics of the colloidal suspension. We start with the fluctuating hydrodynamics framework of Landau and Lifshitz \cite{landau2013fluid}, as formulated in terms of the velocity field $\bfv(\bfx,t)$, which is subject to the incompressibility condition $\bnabla \cdot \bfv=0$, and the pressure filed $p(\bfx,t)$. This framework has been used to study violation of the fluctuation-response relation in suspensions with spatio-temporally structured temperature field \cite{Golestanian2002}, with a related generalization to quantum fluctuations of vacuum developed in Ref. \cite{Kruger2024}. In the viscous limit (i.e. low Reynolds number and no inertial dynamics), this takes the form of an instantaneous and locally enforced stochastic force balance (i.e. momentum conservation) equation 
\beq\label{eq:stokes}
-\eta \bnabla^2 {\bm v}=-\bnabla p+\bff(\bfx,t)+\bxi(\bfx,t),
\eeq
where $\eta$ is the viscosity of the medium, $\bff(\bfx,t)$ is the body-force density, and $\bxi(\bfx,t)$ is a Gaussian white noise term defined via $\aver{\xi_i(\bfx,t)}=0$ and $\aver{\xi_i(\bfx,t) \xi_j(\bfx',t')}=2 \eta \kT \left(-\delta_{ij} \bnabla^2+\pa{i} \pa{j}\right) \delta^d(\bfx-\bfx') \delta(t-t')$. The presence of the colloidal particles necessitates no-slip boundary conditions between the velocities of the particles and the local fluid velocities, namely, $\bfv(\bfr^{\alpha}(t),t)=\dot{\bfr}^{\alpha}(t)$, and the identification of the forces experienced by each particle as the body-force exerted on the fluid, namely, $\bff(\bfx,t)=\sum_{\alpha} \bff^{\alpha} \delta^d(\bfx-\bfr^{\alpha}(t))$.

Using the general path integral formulation of stochastic dynamics through implementation of constraints via Langrange multiplier fields \cite{ZinnJustin2021}, we can construct the path probability distribution weight for the many-body stochastic dynamics of the colloidal particles with hydrodynamic interactions as follows
\beqa
&&{\mathsf P}_\bxi[\{\bfr^{\alpha}(t)\}]=\int {\mathcal D} {\bm v}  {\mathcal D} p \;{\cal J}_{\rm BC} \prod_\alpha \delta\big\{\bfv(\bfr^{\alpha}(t),t)-\dot{\bfr}^{\alpha}(t)\big\}\nonumber \\
&& \hskip1.0cm \times \,{\cal J}_{\rm St} \,\delta\big\{\eta \bnabla^2 {\bm v}-\bnabla p+\bff+\bxi\big\} \; {\cal J}_{\rm in} \,\delta\big\{\bnabla \cdot \bfv\big\},\label{eq:P-0} 
\eeqa
where the Jacobians ${\cal J}_{\rm BC}$ (associated with the boundary conditions), ${\cal J}_{\rm St}$ (associated with the stochastic Stokes equation), and ${\cal J}_{\rm in}$ (associated with the incompressibility condition) ensure normalization of the probability distribution, namely, $\int \prod_\alpha {\mathcal D} \bfr^\alpha(t) {\mathsf P}_\bxi[\{\bfr^{\alpha}(t)\}]=1$. One readily finds that ${\cal J}_{\rm in}={\rm const}$, ${\cal J}_{\rm St}={\rm const}$, and ${\cal J}_{\rm BC}={\rm const} \times \exp\left(-\Theta(0)\int_t \pa{i}^{\alpha} v_i(\bfr^{\alpha}(t),t)\right)={\rm const}$ \cite{ZinnJustin2021}, due to the incompressibility condition. We rewrite Eq. \eqref{eq:P-0} as
\beqa
&&{\mathsf P}_\bxi[\{\bfr^{\alpha}(t)\}]=\frac{1}{Z_{\bxi}}\int \prod_\alpha {\mathcal D} \bphi^\alpha {\mathcal D} {\bm v}  {\mathcal D} p {\mathcal D} {\bar {\bm v}}  {\mathcal D} {\bar p}\nonumber \\
&& \hskip0.8cm\exp\left\{i \int_{t,\bfx} \left[{\bar {\bm v}} \cdot \Big(\eta \bnabla^2 {\bm v}-\bnabla p+\bff+\bxi\Big)+  {\bar p} \Big(\bnabla \cdot \bfv\Big)\right]\right. \nonumber \\
&&\hskip1.5cm\left.+ i \int_{t}  \bphi^\alpha(t) \cdot \Big(\bfv(\bfr^{\alpha}(t),t)-\dot{\bfr}^{\alpha}(t)\Big)\right\},\label{eq:P-1}
\eeqa
where $Z_{\bxi}$ ensures the normalization, and we have used the notation $\int_{t,\bfx} \equiv \int dt \int d^d \bfx$, etc. We then define the field  $\bfh(\bfx,t)=\sum_{\alpha} \bphi^{\alpha} \delta^d(\bfx-\bfr^{\alpha}(t))$ to put all the relevant terms in the action in Eq. (\ref{eq:P-1}) on the same footing, perform noise averaging of the path probability distribution as follows ${\mathsf P}[\{\bfr^{\alpha}(t)\}]\equiv \aver{{\mathsf P}_\bxi[\{\bfr^{\alpha}(t)\}]}$, go into the Fourier space, and perform Gaussian integrals over the fields ${\bm v}$, $p$, ${\bar {\bm v}}$, and ${\bar p}$. After these steps, we obtain
\beqa
&&{\mathsf P}[\{\bfr^{\alpha}(t)\}]=\frac{1}{Z}\int \prod_\alpha {\mathcal D} \bphi^\alpha \nonumber \\
&& \hskip0.4cm\exp\left\{-\kT \int_{\omega,\bfq}\frac{(\delta_{ij}-{\hat q}_i {\hat q}_j)}{\eta q^2} h_i(-\bfq,-\omega) h_j(\bfq,\omega)\right. \nonumber \\
&&\hskip0.4cm\left.+i \int_{\omega,\bfq}\frac{(\delta_{ij}-{\hat q}_i {\hat q}_j)}{\eta q^2} h_i(-\bfq,-\omega) f_j(\bfq,\omega)-i  \int_{t}  \phi^\alpha_i \dot{r}^{\alpha}_i\right\},\nonumber \\
\label{eq:P-2}
\eeqa
where $f_j(\bfq,\omega)=\sum_{\beta} \int_{t'} e^{-i \omega t'+i \bfq \cdot \bfr^\beta(t')} f^{\beta}_j(t')$, $h_j(\bfq,\omega)=\sum_{\beta} \int_{t'} e^{-i \omega t'+i \bfq \cdot \bfr^\beta(t')} \phi^{\beta}_j(t')$, and we have used the shorthand $\int_{\omega,\bfq} \equiv \int \frac{d \omega}{2 \pi} \int \frac{d^d \bfq}{(2 \pi)^d}$.  Here again, the pre-factor $Z$ will ensure normalization of the probability distribution, namely, $\int \prod_\alpha {\mathcal D} \bfr^\alpha(t) {\mathsf P}[\{\bfr^{\alpha}(t)\}]=1$.

Equation (\ref{eq:P-2}) can be written explicitly as a Gaussian integral in terms of the Lagrange multiplier fields $\bphi^\alpha$ as follows
\beqa
&&{\mathsf P}[\{\bfr^{\alpha}(t)\}]=\int \prod_\alpha {\mathcal D} \bphi^\alpha \exp\left\{-\kT  \int_t \cM^{\alpha \beta}_{i j}  \phi^\alpha_i(t)  \phi^\beta_j (t)\right.\nonumber \\ 
&&\hskip1.0cm\left.-i  \int_{t}  \phi^\alpha_i(t) \left[\dot{r}^{\alpha}_i(t)-\cM^{\alpha \beta}_{i j}  f^{\beta}_j(t)\right]\right\}/Z,
\label{eq:P-3}
\eeqa
where $\bcM(\{\bfr^{\alpha}(t)\})$ is the mobility tensor of the colloidal suspension. As an illustration, if we consider spherical colloids of radius $a$ in $d=3$, in the far-field (dilute) limit and free space (no boundaries), the mobility tensor $\bcM(\{\bfr^{\alpha}(t)\})$ is obtained as follows
\beq
\cM^{\alpha \beta}_{i j}(\bfr^{\alpha},\bfr^{\beta})=\left\{
\begin{array}{ll}\ds
  {\delta_{i j}}/{(6 \pi \eta a)}, \quad \quad &  \quad \alpha=\beta, \\
  \cG_{i j}(\bfr^{\alpha}-\bfr^{\beta}),\quad \quad & \quad \alpha \neq \beta.
\end{array}
\right. \label{eq:mobility-hydr-1}
\eeq
Here, the Oseen tensor \cite{oseen1927neuere}
\beq
\cG_{i j}(\bfr)=\frac{1}{8 \pi \eta r} \left[\delta_{i j}+\frac{r_i r_j}{r^2}\right],\label{eq:oseen-hydr-1}
\eeq
is the Green's function (in free space) for Stokes equation of hydrodynamics, describing the velocity profile of viscous and incompressible fluid flow generated by a point force (at the origin). If we have a system with more complex geometric features and confining boundaries, then the elements of the mobility tensor will constitute the appropriate expressions for the friction coefficients as well as the corresponding Green's functions of the Stokes equation with suitable boundary conditions. Note that $\pa{i} \cG_{i j}=0$ due to the incompressibility constraint.

Finally, integrating over the $\bphi^\alpha$ Lagrange multiplier fields in Eq. (\ref{eq:P-3}) yields the closed-form expression as follows 
\beqa
&&{\mathsf P}[\{\bfr^{\alpha}(t)\}]=\frac{1}{Z}\Big(\det \cM^{\alpha \beta}_{i j} \Big)^{-1/2}\nonumber \\ 
&&\times \exp\left\{-\frac{\beta}{4} \int_t \cZ^{\alpha \beta}_{i j} \Big(\dot{r}^{\alpha}_i-\cM^{\alpha \gamma}_{i k}  f^{\gamma}_k\Big)  \Big(\dot{r}^{\beta}_j-\cM^{\beta \delta}_{j l}  f^{\delta}_l\Big)\right\},\nonumber \\
\label{eq:P-4}
\eeqa
where $\beta=1/(\kT)$ and $\bcZ(\{\bfr^{\alpha}(t)\})=\bcM^{-1}$ is the friction tensor. To verify the normalization of Eq. \eqref{eq:P-4} in its natural form $\int \prod_\alpha {\mathcal D} \bfr^\alpha(t) {\mathsf P}[\{\bfr^{\alpha}(t)\}]=1$ and the role played by the measure factor $(\det \bcM)^{-1/2}$, we can introduce a square-root tensor $\bcY$ defined as $\cZ^{\alpha \beta}_{i j}(\{\bfr^{\nu}(t)\})=\cY^{\alpha \gamma}_{i k}(\{\bfr^{\nu}(t)\})\cY^{\beta \gamma}_{j k}(\{\bfr^{\nu}(t)\})$ and a new coordinate system defined via $dR^{\gamma}_{k}=dr^{\alpha}_{i} \cY^{\alpha \gamma}_{i k}(\{\bfr^{\nu}(t)\})$. We then see $\prod_\alpha {\mathcal D} \bfr^\alpha (\det \bcM)^{-1/2}=\prod_\alpha {\mathcal D} \bfr^\alpha (\det \bcY)=\prod_\alpha {\mathcal D} \bfR^\alpha$ and $\cS_{\rm OM}=\frac{1}{4} \int_t \Big(\dot{R}^{\gamma}_k-\cM^{\alpha \rho}_{i m}  f^{\rho}_m \cY^{\alpha \gamma}_{i k}\Big)  \Big(\dot{R}^{\gamma}_k-\cM^{\beta \delta}_{j l}  f^{\delta}_l \cY^{\beta \gamma}_{j k}\Big)$, which is normalized in the usual sense.

\textit{Appendix B: Details of the example of two spheres.---}Here we present some of the details of the calculations reported in the main text. The mobility matrix is defined as
\beq
\cM(x)=\frac{1}{\zeta} 
\begin{bmatrix}
\mu(x) & \lambda(x) \\
\lambda(x) & \mu(x) 
\end{bmatrix},
\eeq
where $\zeta=6 \pi \eta a$. With this definition, it is straightforward to derive Eq. \eqref{eq:P-x-X}. After performing the averages over $X(t)$, we obtain the following expressions
\beq
T \dot{\sigma}=\frac{2  F^2}{\zeta} \aver{\mu+\lambda}+\aver{\dot{x}(t) f(x(t))},
\eeq
and
\beqa
&&\cR^{\alpha \alpha}_{g} (t,t')=\delta(t-t')+\frac{\beta \zeta}{4} \aver{\frac{\dot{x}(t) \dot{x}(t')}{\left[\mu(x(t))-\lambda(x(t))\right]}} \nonumber \\
&&\hskip1.2cm-\frac{\beta}{2} \aver{\dot{x}(t)\frac{\left[\mu(x(t'))-\lambda(x(t'))\right]}{\left[\mu(x(t))-\lambda(x(t))\right]} f(x(t'))},
\eeqa
and
\beqa
\cC^{\alpha \alpha}_{g} (t,t')&=&2 \kT\delta(t-t')+\frac{2  F^2}{\zeta} \aver{\mu+\lambda}\nonumber \\
&&+\frac{\zeta}{2} \aver{\frac{\dot{x}(t) \dot{x}(t')}{\left[\mu(x(t))-\lambda(x(t))\right]}},
\eeqa
where the averages are to be performed using the following marginal distribution
\beq
P[x(t)]=\frac{1}{Z} \prod_t \frac{1}{\sqrt{\mu(x(t))-\lambda(x(t))}} \, e^{-\beta S_{\rm OM}},\label{eq:P-x}
\eeq
where 
\beq
S_{\rm OM}[x(t)]=\frac{\zeta}{8}\int_t \frac{\left[\dot{x}(t)-\frac{2}{\zeta}\left(\mu(x(t))-\lambda(x(t))\right) f(x(t))\right]^2}{\left[\mu(x(t))-\lambda(x(t))\right]}.
\eeq
The coordinate transformation can be readily implemented, resulting in the Gaussian weight reported in the main text above.

\acknowledgements
I acknowledge support from the Max Planck School Matter to Life and the MaxSynBio Consortium which are jointly funded by the Federal Ministry of Education and Research (BMBF) of Germany and the Max Planck Society.

\bibliography{bib_stoch.bib}

\begin{thebibliography}{56}%
\makeatletter
\providecommand \@ifxundefined [1]{%
 \@ifx{#1\undefined}
}%
\providecommand \@ifnum [1]{%
 \ifnum #1\expandafter \@firstoftwo
 \else \expandafter \@secondoftwo
 \fi
}%
\providecommand \@ifx [1]{%
 \ifx #1\expandafter \@firstoftwo
 \else \expandafter \@secondoftwo
 \fi
}%
\providecommand \natexlab [1]{#1}%
\providecommand \enquote  [1]{``#1''}%
\providecommand \bibnamefont  [1]{#1}%
\providecommand \bibfnamefont [1]{#1}%
\providecommand \citenamefont [1]{#1}%
\providecommand \href@noop [0]{\@secondoftwo}%
\providecommand \href [0]{\begingroup \@sanitize@url \@href}%
\providecommand \@href[1]{\@@startlink{#1}\@@href}%
\providecommand \@@href[1]{\endgroup#1\@@endlink}%
\providecommand \@sanitize@url [0]{\catcode `\\12\catcode `\$12\catcode
  `\&12\catcode `\#12\catcode `\^12\catcode `\_12\catcode `\%12\relax}%
\providecommand \@@startlink[1]{}%
\providecommand \@@endlink[0]{}%
\providecommand \url  [0]{\begingroup\@sanitize@url \@url }%
\providecommand \@url [1]{\endgroup\@href {#1}{\urlprefix }}%
\providecommand \urlprefix  [0]{URL }%
\providecommand \Eprint [0]{\href }%
\providecommand \doibase [0]{https://doi.org/}%
\providecommand \selectlanguage [0]{\@gobble}%
\providecommand \bibinfo  [0]{\@secondoftwo}%
\providecommand \bibfield  [0]{\@secondoftwo}%
\providecommand \translation [1]{[#1]}%
\providecommand \BibitemOpen [0]{}%
\providecommand \bibitemStop [0]{}%
\providecommand \bibitemNoStop [0]{.\EOS\space}%
\providecommand \EOS [0]{\spacefactor3000\relax}%
\providecommand \BibitemShut  [1]{\csname bibitem#1\endcsname}%
\let\auto@bib@innerbib\@empty
\bibitem [{\citenamefont {Onsager}(1931)}]{onsager2}%
  \BibitemOpen
  \bibfield  {author} {\bibinfo {author} {\bibfnamefont {L.}~\bibnamefont
  {Onsager}},\ }\bibfield  {title} {\bibinfo {title} {Reciprocal relations in
  irreversible processes. ii.},\ }\href
  {https://doi.org/10.1103/PhysRev.38.2265} {\bibfield  {journal} {\bibinfo
  {journal} {Phys. Rev.}\ }\textbf {\bibinfo {volume} {38}},\ \bibinfo {pages}
  {2265} (\bibinfo {year} {1931})}\BibitemShut {NoStop}%
\bibitem [{\citenamefont {Cugliandolo}\ \emph
  {et~al.}(1997{\natexlab{a}})\citenamefont {Cugliandolo}, \citenamefont
  {Kurchan},\ and\ \citenamefont {Peliti}}]{LetPeliti1997}%
  \BibitemOpen
  \bibfield  {author} {\bibinfo {author} {\bibfnamefont {L.~F.}\ \bibnamefont
  {Cugliandolo}}, \bibinfo {author} {\bibfnamefont {J.}~\bibnamefont
  {Kurchan}},\ and\ \bibinfo {author} {\bibfnamefont {L.}~\bibnamefont
  {Peliti}},\ }\bibfield  {title} {\bibinfo {title} {Energy flow, partial
  equilibration, and effective temperatures in systems with slow dynamics},\
  }\href {https://doi.org/10.1103/PhysRevE.55.3898} {\bibfield  {journal}
  {\bibinfo  {journal} {Phys. Rev. E}\ }\textbf {\bibinfo {volume} {55}},\
  \bibinfo {pages} {3898} (\bibinfo {year} {1997}{\natexlab{a}})}\BibitemShut
  {NoStop}%
\bibitem [{\citenamefont {Cugliandolo}\ \emph
  {et~al.}(1997{\natexlab{b}})\citenamefont {Cugliandolo}, \citenamefont
  {Dean},\ and\ \citenamefont {Kurchan}}]{LetDean1997}%
  \BibitemOpen
  \bibfield  {author} {\bibinfo {author} {\bibfnamefont {L.~F.}\ \bibnamefont
  {Cugliandolo}}, \bibinfo {author} {\bibfnamefont {D.~S.}\ \bibnamefont
  {Dean}},\ and\ \bibinfo {author} {\bibfnamefont {J.}~\bibnamefont
  {Kurchan}},\ }\bibfield  {title} {\bibinfo {title} {Fluctuation-dissipation
  theorems and entropy production in relaxational systems},\ }\href
  {https://doi.org/10.1103/PhysRevLett.79.2168} {\bibfield  {journal} {\bibinfo
   {journal} {Phys. Rev. Lett.}\ }\textbf {\bibinfo {volume} {79}},\ \bibinfo
  {pages} {2168} (\bibinfo {year} {1997}{\natexlab{b}})}\BibitemShut {NoStop}%
\bibitem [{\citenamefont {Golestanian}\ and\ \citenamefont
  {Ajdari}(2002)}]{Golestanian2002}%
  \BibitemOpen
  \bibfield  {author} {\bibinfo {author} {\bibfnamefont {R.}~\bibnamefont
  {Golestanian}}\ and\ \bibinfo {author} {\bibfnamefont {A.}~\bibnamefont
  {Ajdari}},\ }\bibfield  {title} {\bibinfo {title} {Tracer diffusivity in a
  time- or space-dependent temperature field},\ }\href
  {https://doi.org/10.1209/epl/i2002-00113-x} {\bibfield  {journal} {\bibinfo
  {journal} {Europhys. Lett.}\ }\textbf {\bibinfo {volume} {59}},\ \bibinfo
  {pages} {800} (\bibinfo {year} {2002})}\BibitemShut {NoStop}%
\bibitem [{\citenamefont {Crisanti}\ and\ \citenamefont
  {Ritort}(2003)}]{Crisanti2003}%
  \BibitemOpen
  \bibfield  {author} {\bibinfo {author} {\bibfnamefont {A.}~\bibnamefont
  {Crisanti}}\ and\ \bibinfo {author} {\bibfnamefont {F.}~\bibnamefont
  {Ritort}},\ }\bibfield  {title} {\bibinfo {title} {Violation of the
  fluctuation–dissipation theorem in glassy systems: basic notions and the
  numerical evidence},\ }\href {https://doi.org/10.1088/0305-4470/36/21/201}
  {\bibfield  {journal} {\bibinfo  {journal} {Journal of Physics A:
  Mathematical and General}\ }\textbf {\bibinfo {volume} {36}},\ \bibinfo
  {pages} {R181–R290} (\bibinfo {year} {2003})}\BibitemShut {NoStop}%
\bibitem [{\citenamefont {Harada}\ and\ \citenamefont
  {Sasa}(2005)}]{haradaEqualityConnectingEnergy2005}%
  \BibitemOpen
  \bibfield  {author} {\bibinfo {author} {\bibfnamefont {T.}~\bibnamefont
  {Harada}}\ and\ \bibinfo {author} {\bibfnamefont {S.-i.}\ \bibnamefont
  {Sasa}},\ }\bibfield  {title} {\bibinfo {title} {Equality {{Connecting Energy
  Dissipation}} with a {{Violation}} of the {{Fluctuation-Response
  Relation}}},\ }\href {https://doi.org/10.1103/PhysRevLett.95.130602}
  {\bibfield  {journal} {\bibinfo  {journal} {Phys. Rev. Lett.}\ }\textbf
  {\bibinfo {volume} {95}},\ \bibinfo {pages} {130602} (\bibinfo {year}
  {2005})}\BibitemShut {NoStop}%
\bibitem [{\citenamefont {Harada}\ and\ \citenamefont
  {Sasa}(2006)}]{haradaEnergyDissipationViolation2006}%
  \BibitemOpen
  \bibfield  {author} {\bibinfo {author} {\bibfnamefont {T.}~\bibnamefont
  {Harada}}\ and\ \bibinfo {author} {\bibfnamefont {S.-i.}\ \bibnamefont
  {Sasa}},\ }\bibfield  {title} {\bibinfo {title} {Energy dissipation and
  violation of the fluctuation-response relation in nonequilibrium {{Langevin}}
  systems},\ }\href {https://doi.org/10.1103/PhysRevE.73.026131} {\bibfield
  {journal} {\bibinfo  {journal} {Phys. Rev. E}\ }\textbf {\bibinfo {volume}
  {73}},\ \bibinfo {pages} {026131} (\bibinfo {year} {2006})}\BibitemShut
  {NoStop}%
\bibitem [{\citenamefont {Speck}\ and\ \citenamefont
  {Seifert}(2006)}]{Speck2006}%
  \BibitemOpen
  \bibfield  {author} {\bibinfo {author} {\bibfnamefont {T.}~\bibnamefont
  {Speck}}\ and\ \bibinfo {author} {\bibfnamefont {U.}~\bibnamefont
  {Seifert}},\ }\bibfield  {title} {\bibinfo {title} {Restoring a
  fluctuation-dissipation theorem in a nonequilibrium steady state},\ }\href
  {https://doi.org/10.1209/epl/i2005-10549-4} {\bibfield  {journal} {\bibinfo
  {journal} {Europhysics Letters (EPL)}\ }\textbf {\bibinfo {volume} {74}},\
  \bibinfo {pages} {391–396} (\bibinfo {year} {2006})}\BibitemShut {NoStop}%
\bibitem [{\citenamefont {Baiesi}\ \emph {et~al.}(2009)\citenamefont {Baiesi},
  \citenamefont {Maes},\ and\ \citenamefont {Wynants}}]{Maes2009}%
  \BibitemOpen
  \bibfield  {author} {\bibinfo {author} {\bibfnamefont {M.}~\bibnamefont
  {Baiesi}}, \bibinfo {author} {\bibfnamefont {C.}~\bibnamefont {Maes}},\ and\
  \bibinfo {author} {\bibfnamefont {B.}~\bibnamefont {Wynants}},\ }\bibfield
  {title} {\bibinfo {title} {Fluctuations and response of nonequilibrium
  states},\ }\href {https://doi.org/10.1103/PhysRevLett.103.010602} {\bibfield
  {journal} {\bibinfo  {journal} {Phys. Rev. Lett.}\ }\textbf {\bibinfo
  {volume} {103}},\ \bibinfo {pages} {010602} (\bibinfo {year}
  {2009})}\BibitemShut {NoStop}%
\bibitem [{\citenamefont {Prost}\ \emph {et~al.}(2009)\citenamefont {Prost},
  \citenamefont {Joanny},\ and\ \citenamefont {Parrondo}}]{Parrondo2009}%
  \BibitemOpen
  \bibfield  {author} {\bibinfo {author} {\bibfnamefont {J.}~\bibnamefont
  {Prost}}, \bibinfo {author} {\bibfnamefont {J.-F.}\ \bibnamefont {Joanny}},\
  and\ \bibinfo {author} {\bibfnamefont {J.~M.~R.}\ \bibnamefont {Parrondo}},\
  }\bibfield  {title} {\bibinfo {title} {Generalized fluctuation-dissipation
  theorem for steady-state systems},\ }\href
  {https://doi.org/10.1103/PhysRevLett.103.090601} {\bibfield  {journal}
  {\bibinfo  {journal} {Phys. Rev. Lett.}\ }\textbf {\bibinfo {volume} {103}},\
  \bibinfo {pages} {090601} (\bibinfo {year} {2009})}\BibitemShut {NoStop}%
\bibitem [{\citenamefont {Harada}(2009)}]{Harada2009}%
  \BibitemOpen
  \bibfield  {author} {\bibinfo {author} {\bibfnamefont {T.}~\bibnamefont
  {Harada}},\ }\bibfield  {title} {\bibinfo {title} {Macroscopic expression
  connecting the rate of energy dissipation with the violation of the
  fluctuation response relation},\ }\href
  {https://doi.org/10.1103/PhysRevE.79.030106} {\bibfield  {journal} {\bibinfo
  {journal} {Phys. Rev. E}\ }\textbf {\bibinfo {volume} {79}},\ \bibinfo
  {pages} {030106} (\bibinfo {year} {2009})}\BibitemShut {NoStop}%
\bibitem [{\citenamefont {Turlier}\ \emph {et~al.}(2016)\citenamefont
  {Turlier}, \citenamefont {Fedosov}, \citenamefont {Audoly}, \citenamefont
  {Auth}, \citenamefont {Gov}, \citenamefont {Sykes}, \citenamefont {Joanny},
  \citenamefont {Gompper},\ and\ \citenamefont {Betz}}]{Turlier2016}%
  \BibitemOpen
  \bibfield  {author} {\bibinfo {author} {\bibfnamefont {H.}~\bibnamefont
  {Turlier}}, \bibinfo {author} {\bibfnamefont {D.~A.}\ \bibnamefont
  {Fedosov}}, \bibinfo {author} {\bibfnamefont {B.}~\bibnamefont {Audoly}},
  \bibinfo {author} {\bibfnamefont {T.}~\bibnamefont {Auth}}, \bibinfo {author}
  {\bibfnamefont {N.~S.}\ \bibnamefont {Gov}}, \bibinfo {author} {\bibfnamefont
  {C.}~\bibnamefont {Sykes}}, \bibinfo {author} {\bibfnamefont {J.-F.}\
  \bibnamefont {Joanny}}, \bibinfo {author} {\bibfnamefont {G.}~\bibnamefont
  {Gompper}},\ and\ \bibinfo {author} {\bibfnamefont {T.}~\bibnamefont
  {Betz}},\ }\bibfield  {title} {\bibinfo {title} {Equilibrium physics
  breakdown reveals the active nature of red blood cell flickering},\ }\href
  {https://doi.org/10.1038/nphys3621} {\bibfield  {journal} {\bibinfo
  {journal} {Nature Physics}\ }\textbf {\bibinfo {volume} {12}},\ \bibinfo
  {pages} {513–519} (\bibinfo {year} {2016})}\BibitemShut {NoStop}%
\bibitem [{\citenamefont {Abah}\ and\ \citenamefont {Lutz}(2016)}]{Abah2016}%
  \BibitemOpen
  \bibfield  {author} {\bibinfo {author} {\bibfnamefont {O.}~\bibnamefont
  {Abah}}\ and\ \bibinfo {author} {\bibfnamefont {E.}~\bibnamefont {Lutz}},\
  }\bibfield  {title} {\bibinfo {title} {Optimal performance of a quantum otto
  refrigerator},\ }\href {https://doi.org/10.1209/0295-5075/113/60002}
  {\bibfield  {journal} {\bibinfo  {journal} {EPL (Europhysics Letters)}\
  }\textbf {\bibinfo {volume} {113}},\ \bibinfo {pages} {60002} (\bibinfo
  {year} {2016})}\BibitemShut {NoStop}%
\bibitem [{\citenamefont {Maes}(2020)}]{Maes2020}%
  \BibitemOpen
  \bibfield  {author} {\bibinfo {author} {\bibfnamefont {C.}~\bibnamefont
  {Maes}},\ }\bibfield  {title} {\bibinfo {title} {Response theory: A
  trajectory-based approach},\ }\bibfield  {journal} {\bibinfo  {journal}
  {Frontiers in Physics}\ }\textbf {\bibinfo {volume} {8}},\ \href
  {https://doi.org/10.3389/fphy.2020.00229} {10.3389/fphy.2020.00229} (\bibinfo
  {year} {2020})\BibitemShut {NoStop}%
\bibitem [{\citenamefont {Gompper}\ \emph {et~al.}(2020)\citenamefont
  {Gompper}, \citenamefont {Winkler}, \citenamefont {Speck}, \citenamefont
  {Solon}, \citenamefont {Nardini}, \citenamefont {Peruani}, \citenamefont
  {L{\"o}wen}, \citenamefont {Golestanian}, \citenamefont {Kaupp},
  \citenamefont {Alvarez} \emph {et~al.}}]{Gompper2020}%
  \BibitemOpen
  \bibfield  {author} {\bibinfo {author} {\bibfnamefont {G.}~\bibnamefont
  {Gompper}}, \bibinfo {author} {\bibfnamefont {R.~G.}\ \bibnamefont
  {Winkler}}, \bibinfo {author} {\bibfnamefont {T.}~\bibnamefont {Speck}},
  \bibinfo {author} {\bibfnamefont {A.}~\bibnamefont {Solon}}, \bibinfo
  {author} {\bibfnamefont {C.}~\bibnamefont {Nardini}}, \bibinfo {author}
  {\bibfnamefont {F.}~\bibnamefont {Peruani}}, \bibinfo {author} {\bibfnamefont
  {H.}~\bibnamefont {L{\"o}wen}}, \bibinfo {author} {\bibfnamefont
  {R.}~\bibnamefont {Golestanian}}, \bibinfo {author} {\bibfnamefont {U.~B.}\
  \bibnamefont {Kaupp}}, \bibinfo {author} {\bibfnamefont {L.}~\bibnamefont
  {Alvarez}}, \emph {et~al.},\ }\bibfield  {title} {\bibinfo {title} {The 2020
  motile active matter roadmap},\ }\href
  {https://doi.org/10.1088/1361-648X/ab6348} {\bibfield  {journal} {\bibinfo
  {journal} {J. Phys.: Condens. Matter}\ }\textbf {\bibinfo {volume} {32}},\
  \bibinfo {pages} {193001} (\bibinfo {year} {2020})}\BibitemShut {NoStop}%
\bibitem [{\citenamefont {Battle}\ \emph {et~al.}(2016)\citenamefont {Battle},
  \citenamefont {Broedersz}, \citenamefont {Fakhri}, \citenamefont {Geyer},
  \citenamefont {Howard}, \citenamefont {Schmidt},\ and\ \citenamefont
  {MacKintosh}}]{battle2016broken}%
  \BibitemOpen
  \bibfield  {author} {\bibinfo {author} {\bibfnamefont {C.}~\bibnamefont
  {Battle}}, \bibinfo {author} {\bibfnamefont {C.~P.}\ \bibnamefont
  {Broedersz}}, \bibinfo {author} {\bibfnamefont {N.}~\bibnamefont {Fakhri}},
  \bibinfo {author} {\bibfnamefont {V.~F.}\ \bibnamefont {Geyer}}, \bibinfo
  {author} {\bibfnamefont {J.}~\bibnamefont {Howard}}, \bibinfo {author}
  {\bibfnamefont {C.~F.}\ \bibnamefont {Schmidt}},\ and\ \bibinfo {author}
  {\bibfnamefont {F.~C.}\ \bibnamefont {MacKintosh}},\ }\bibfield  {title}
  {\bibinfo {title} {Broken detailed balance at mesoscopic scales in active
  biological systems},\ }\href {https://doi.org/10.1126/science.aac8167}
  {\bibfield  {journal} {\bibinfo  {journal} {Science}\ }\textbf {\bibinfo
  {volume} {352}},\ \bibinfo {pages} {604} (\bibinfo {year}
  {2016})}\BibitemShut {NoStop}%
\bibitem [{\citenamefont {Nardini}\ \emph {et~al.}(2017)\citenamefont
  {Nardini}, \citenamefont {Fodor}, \citenamefont {Tjhung}, \citenamefont {van
  Wijland}, \citenamefont {Tailleur},\ and\ \citenamefont
  {Cates}}]{nardini2017entropy}%
  \BibitemOpen
  \bibfield  {author} {\bibinfo {author} {\bibfnamefont {C.}~\bibnamefont
  {Nardini}}, \bibinfo {author} {\bibfnamefont {E.}~\bibnamefont {Fodor}},
  \bibinfo {author} {\bibfnamefont {E.}~\bibnamefont {Tjhung}}, \bibinfo
  {author} {\bibfnamefont {F.}~\bibnamefont {van Wijland}}, \bibinfo {author}
  {\bibfnamefont {J.}~\bibnamefont {Tailleur}},\ and\ \bibinfo {author}
  {\bibfnamefont {M.~E.}\ \bibnamefont {Cates}},\ }\bibfield  {title} {\bibinfo
  {title} {Entropy production in field theories without time-reversal symmetry:
  Quantifying the non-equilibrium character of active matter},\ }\href
  {https://doi.org/10.1103/PhysRevX.7.021007} {\bibfield  {journal} {\bibinfo
  {journal} {Phys. Rev. X}\ }\textbf {\bibinfo {volume} {7}},\ \bibinfo {pages}
  {021007} (\bibinfo {year} {2017})}\BibitemShut {NoStop}%
\bibitem [{\citenamefont {Pietzonka}\ and\ \citenamefont
  {Seifert}(2017)}]{pietzonka2017entropy}%
  \BibitemOpen
  \bibfield  {author} {\bibinfo {author} {\bibfnamefont {P.}~\bibnamefont
  {Pietzonka}}\ and\ \bibinfo {author} {\bibfnamefont {U.}~\bibnamefont
  {Seifert}},\ }\bibfield  {title} {\bibinfo {title} {Entropy production of
  active particles and for particles in active baths},\ }\href
  {https://doi.org/10.1088/1751-8121/aa91b9} {\bibfield  {journal} {\bibinfo
  {journal} {J. Phys. A: Math}\ }\textbf {\bibinfo {volume} {51}},\ \bibinfo
  {pages} {01LT01} (\bibinfo {year} {2017})}\BibitemShut {NoStop}%
\bibitem [{\citenamefont {Shankar}\ and\ \citenamefont
  {Marchetti}(2018)}]{shankar2018hidden}%
  \BibitemOpen
  \bibfield  {author} {\bibinfo {author} {\bibfnamefont {S.}~\bibnamefont
  {Shankar}}\ and\ \bibinfo {author} {\bibfnamefont {M.~C.}\ \bibnamefont
  {Marchetti}},\ }\bibfield  {title} {\bibinfo {title} {Hidden entropy
  production and work fluctuations in an ideal active gas},\ }\href
  {https://doi.org/10.1103/PhysRevE.98.020604} {\bibfield  {journal} {\bibinfo
  {journal} {Phys. Rev. E}\ }\textbf {\bibinfo {volume} {98}},\ \bibinfo
  {pages} {020604} (\bibinfo {year} {2018})}\BibitemShut {NoStop}%
\bibitem [{\citenamefont {J{\"u}licher}\ \emph {et~al.}(1997)\citenamefont
  {J{\"u}licher}, \citenamefont {Ajdari},\ and\ \citenamefont
  {Prost}}]{julicher1997modeling}%
  \BibitemOpen
  \bibfield  {author} {\bibinfo {author} {\bibfnamefont {F.}~\bibnamefont
  {J{\"u}licher}}, \bibinfo {author} {\bibfnamefont {A.}~\bibnamefont
  {Ajdari}},\ and\ \bibinfo {author} {\bibfnamefont {J.}~\bibnamefont
  {Prost}},\ }\bibfield  {title} {\bibinfo {title} {Modeling molecular
  motors},\ }\href@noop {} {\bibfield  {journal} {\bibinfo  {journal} {Rev.
  Mod. Phys.}\ }\textbf {\bibinfo {volume} {69}},\ \bibinfo {pages} {1269}
  (\bibinfo {year} {1997})}\BibitemShut {NoStop}%
\bibitem [{\citenamefont {Ramaswamy}(2010)}]{ramaswamy2010}%
  \BibitemOpen
  \bibfield  {author} {\bibinfo {author} {\bibfnamefont {S.}~\bibnamefont
  {Ramaswamy}},\ }\bibfield  {title} {\bibinfo {title} {The mechanics and
  statistics of active matter},\ }\href
  {https://doi.org/10.1146/annurev-conmatphys-070909-104101} {\bibfield
  {journal} {\bibinfo  {journal} {Annual Review of Condensed Matter Physics}\
  }\textbf {\bibinfo {volume} {1}},\ \bibinfo {pages} {323–345} (\bibinfo
  {year} {2010})}\BibitemShut {NoStop}%
\bibitem [{\citenamefont {Mugnai}\ \emph {et~al.}(2020)\citenamefont {Mugnai},
  \citenamefont {Hyeon}, \citenamefont {Hinczewski},\ and\ \citenamefont
  {Thirumalai}}]{mugnai2020theoretical}%
  \BibitemOpen
  \bibfield  {author} {\bibinfo {author} {\bibfnamefont {M.~L.}\ \bibnamefont
  {Mugnai}}, \bibinfo {author} {\bibfnamefont {C.}~\bibnamefont {Hyeon}},
  \bibinfo {author} {\bibfnamefont {M.}~\bibnamefont {Hinczewski}},\ and\
  \bibinfo {author} {\bibfnamefont {D.}~\bibnamefont {Thirumalai}},\ }\bibfield
   {title} {\bibinfo {title} {Theoretical perspectives on biological
  machines},\ }\href {https://doi.org/10.1103/RevModPhys.92.025001} {\bibfield
  {journal} {\bibinfo  {journal} {Rev. Mod. Phys.}\ }\textbf {\bibinfo {volume}
  {92}},\ \bibinfo {pages} {025001} (\bibinfo {year} {2020})}\BibitemShut
  {NoStop}%
\bibitem [{\citenamefont {Borsley}\ \emph {et~al.}(2022)\citenamefont
  {Borsley}, \citenamefont {Leigh},\ and\ \citenamefont
  {Roberts}}]{borsley2022chemical}%
  \BibitemOpen
  \bibfield  {author} {\bibinfo {author} {\bibfnamefont {S.}~\bibnamefont
  {Borsley}}, \bibinfo {author} {\bibfnamefont {D.~A.}\ \bibnamefont {Leigh}},\
  and\ \bibinfo {author} {\bibfnamefont {B.~M.~W.}\ \bibnamefont {Roberts}},\
  }\bibfield  {title} {\bibinfo {title} {Chemical fuels for molecular
  machinery},\ }\href {https://doi.org/10.1038/s41557-022-00970-9} {\bibfield
  {journal} {\bibinfo  {journal} {Nat. Chem.}\ }\textbf {\bibinfo {volume}
  {14}},\ \bibinfo {pages} {728} (\bibinfo {year} {2022})}\BibitemShut
  {NoStop}%
\bibitem [{\citenamefont {Pumm}\ \emph {et~al.}(2022)\citenamefont {Pumm},
  \citenamefont {Engelen}, \citenamefont {Kopperger}, \citenamefont {Isensee},
  \citenamefont {Vogt}, \citenamefont {Kozina}, \citenamefont {Kube},
  \citenamefont {Honemann}, \citenamefont {Bertosin}, \citenamefont
  {Langecker}, \citenamefont {Golestanian}, \citenamefont {Simmel},\ and\
  \citenamefont {Dietz}}]{Pumm2022}%
  \BibitemOpen
  \bibfield  {author} {\bibinfo {author} {\bibfnamefont {A.-K.}\ \bibnamefont
  {Pumm}}, \bibinfo {author} {\bibfnamefont {W.}~\bibnamefont {Engelen}},
  \bibinfo {author} {\bibfnamefont {E.}~\bibnamefont {Kopperger}}, \bibinfo
  {author} {\bibfnamefont {J.}~\bibnamefont {Isensee}}, \bibinfo {author}
  {\bibfnamefont {M.}~\bibnamefont {Vogt}}, \bibinfo {author} {\bibfnamefont
  {V.}~\bibnamefont {Kozina}}, \bibinfo {author} {\bibfnamefont
  {M.}~\bibnamefont {Kube}}, \bibinfo {author} {\bibfnamefont {M.~N.}\
  \bibnamefont {Honemann}}, \bibinfo {author} {\bibfnamefont {E.}~\bibnamefont
  {Bertosin}}, \bibinfo {author} {\bibfnamefont {M.}~\bibnamefont {Langecker}},
  \bibinfo {author} {\bibfnamefont {R.}~\bibnamefont {Golestanian}}, \bibinfo
  {author} {\bibfnamefont {F.~C.}\ \bibnamefont {Simmel}},\ and\ \bibinfo
  {author} {\bibfnamefont {H.}~\bibnamefont {Dietz}},\ }\bibfield  {title}
  {\bibinfo {title} {A {DNA} origami rotary ratchet motor},\ }\href
  {https://doi.org/10.1038/s41586-022-04910-y} {\bibfield  {journal} {\bibinfo
  {journal} {Nature}\ }\textbf {\bibinfo {volume} {607}},\ \bibinfo {pages}
  {492–498} (\bibinfo {year} {2022})}\BibitemShut {NoStop}%
\bibitem [{\citenamefont {Shi}\ \emph {et~al.}(2022)\citenamefont {Shi},
  \citenamefont {Pumm}, \citenamefont {Isensee}, \citenamefont {Zhao},
  \citenamefont {Verschueren}, \citenamefont {Martin-Gonzalez}, \citenamefont
  {Golestanian}, \citenamefont {Dietz},\ and\ \citenamefont
  {Dekker}}]{Shi2022}%
  \BibitemOpen
  \bibfield  {author} {\bibinfo {author} {\bibfnamefont {X.}~\bibnamefont
  {Shi}}, \bibinfo {author} {\bibfnamefont {A.-K.}\ \bibnamefont {Pumm}},
  \bibinfo {author} {\bibfnamefont {J.}~\bibnamefont {Isensee}}, \bibinfo
  {author} {\bibfnamefont {W.}~\bibnamefont {Zhao}}, \bibinfo {author}
  {\bibfnamefont {D.}~\bibnamefont {Verschueren}}, \bibinfo {author}
  {\bibfnamefont {A.}~\bibnamefont {Martin-Gonzalez}}, \bibinfo {author}
  {\bibfnamefont {R.}~\bibnamefont {Golestanian}}, \bibinfo {author}
  {\bibfnamefont {H.}~\bibnamefont {Dietz}},\ and\ \bibinfo {author}
  {\bibfnamefont {C.}~\bibnamefont {Dekker}},\ }\bibfield  {title} {\bibinfo
  {title} {Sustained unidirectional rotation of a self-organized {DNA} rotor on
  a nanopore},\ }\href {https://doi.org/10.1038/s41567-022-01683-z} {\bibfield
  {journal} {\bibinfo  {journal} {Nat. Phys.}\ }\textbf {\bibinfo {volume}
  {18}},\ \bibinfo {pages} {1105–1111} (\bibinfo {year} {2022})}\BibitemShut
  {NoStop}%
\bibitem [{\citenamefont {Shi}\ \emph {et~al.}(2023)\citenamefont {Shi},
  \citenamefont {Pumm}, \citenamefont {Maffeo}, \citenamefont {Kohler},
  \citenamefont {Feigl}, \citenamefont {Zhao}, \citenamefont {Verschueren},
  \citenamefont {Golestanian}, \citenamefont {Aksimentiev}, \citenamefont
  {Dietz},\ and\ \citenamefont {Dekker}}]{Shi2023}%
  \BibitemOpen
  \bibfield  {author} {\bibinfo {author} {\bibfnamefont {X.}~\bibnamefont
  {Shi}}, \bibinfo {author} {\bibfnamefont {A.-K.}\ \bibnamefont {Pumm}},
  \bibinfo {author} {\bibfnamefont {C.}~\bibnamefont {Maffeo}}, \bibinfo
  {author} {\bibfnamefont {F.}~\bibnamefont {Kohler}}, \bibinfo {author}
  {\bibfnamefont {E.}~\bibnamefont {Feigl}}, \bibinfo {author} {\bibfnamefont
  {W.}~\bibnamefont {Zhao}}, \bibinfo {author} {\bibfnamefont {D.}~\bibnamefont
  {Verschueren}}, \bibinfo {author} {\bibfnamefont {R.}~\bibnamefont
  {Golestanian}}, \bibinfo {author} {\bibfnamefont {A.}~\bibnamefont
  {Aksimentiev}}, \bibinfo {author} {\bibfnamefont {H.}~\bibnamefont {Dietz}},\
  and\ \bibinfo {author} {\bibfnamefont {C.}~\bibnamefont {Dekker}},\
  }\bibfield  {title} {\bibinfo {title} {A {DNA} turbine powered by a
  transmembrane potential across a nanopore},\ }\href
  {http://dx.doi.org/10.1038/s41565-023-01527-8} {\bibfield  {journal}
  {\bibinfo  {journal} {Nat. Nanotechnol.}\ } (\bibinfo {year}
  {2023})}\BibitemShut {NoStop}%
\bibitem [{\citenamefont {Golestanian}(2022)}]{Golestanian2019phoretic}%
  \BibitemOpen
  \bibfield  {author} {\bibinfo {author} {\bibfnamefont {R.}~\bibnamefont
  {Golestanian}},\ }\bibfield  {title} {\bibinfo {title} {{Phoretic Active
  Matter}},\ }in\ \href {https://doi.org/10.1093/oso/9780192858313.003.0008}
  {\emph {\bibinfo {booktitle} {{Active Matter and Nonequilibrium Statistical
  Physics: Lecture Notes of the Les Houches Summer School: Volume 112,
  September 2018}}}}\ (\bibinfo  {publisher} {Oxford University Press},\
  \bibinfo {year} {2022})\BibitemShut {NoStop}%
\bibitem [{\citenamefont {Golestanian}\ and\ \citenamefont
  {Ajdari}(2008{\natexlab{a}})}]{RG2008}%
  \BibitemOpen
  \bibfield  {author} {\bibinfo {author} {\bibfnamefont {R.}~\bibnamefont
  {Golestanian}}\ and\ \bibinfo {author} {\bibfnamefont {A.}~\bibnamefont
  {Ajdari}},\ }\bibfield  {title} {\bibinfo {title} {Mechanical response of a
  small swimmer driven by conformational transitions},\ }\href
  {https://doi.org/10.1103/PhysRevLett.100.038101} {\bibfield  {journal}
  {\bibinfo  {journal} {Phys. Rev. Lett.}\ }\textbf {\bibinfo {volume} {100}},\
  \bibinfo {pages} {038101} (\bibinfo {year} {2008}{\natexlab{a}})}\BibitemShut
  {NoStop}%
\bibitem [{\citenamefont {Chatzittofi}\ \emph {et~al.}(2024)\citenamefont
  {Chatzittofi}, \citenamefont {Agudo-Canalejo},\ and\ \citenamefont
  {Golestanian}}]{Mike3SS2024}%
  \BibitemOpen
  \bibfield  {author} {\bibinfo {author} {\bibfnamefont {M.}~\bibnamefont
  {Chatzittofi}}, \bibinfo {author} {\bibfnamefont {J.}~\bibnamefont
  {Agudo-Canalejo}},\ and\ \bibinfo {author} {\bibfnamefont {R.}~\bibnamefont
  {Golestanian}},\ }\bibfield  {title} {\bibinfo {title} {Entropy production
  and thermodynamic inference for stochastic microswimmers},\ }\href
  {https://doi.org/10.1103/PhysRevResearch.6.L022044} {\bibfield  {journal}
  {\bibinfo  {journal} {Phys. Rev. Res.}\ }\textbf {\bibinfo {volume} {6}},\
  \bibinfo {pages} {L022044} (\bibinfo {year} {2024})}\BibitemShut {NoStop}%
\bibitem [{\citenamefont {Nasouri}\ \emph {et~al.}(2021)\citenamefont
  {Nasouri}, \citenamefont {Vilfan},\ and\ \citenamefont
  {Golestanian}}]{Babak2021}%
  \BibitemOpen
  \bibfield  {author} {\bibinfo {author} {\bibfnamefont {B.}~\bibnamefont
  {Nasouri}}, \bibinfo {author} {\bibfnamefont {A.}~\bibnamefont {Vilfan}},\
  and\ \bibinfo {author} {\bibfnamefont {R.}~\bibnamefont {Golestanian}},\
  }\bibfield  {title} {\bibinfo {title} {Minimum dissipation theorem for
  microswimmers},\ }\href {https://doi.org/10.1103/PhysRevLett.126.034503}
  {\bibfield  {journal} {\bibinfo  {journal} {Phys. Rev. Lett.}\ }\textbf
  {\bibinfo {volume} {126}},\ \bibinfo {pages} {034503} (\bibinfo {year}
  {2021})}\BibitemShut {NoStop}%
\bibitem [{\citenamefont {Daddi-Moussa-Ider}\ \emph {et~al.}(2023)\citenamefont
  {Daddi-Moussa-Ider}, \citenamefont {Golestanian},\ and\ \citenamefont
  {Vilfan}}]{daddi2023minimum}%
  \BibitemOpen
  \bibfield  {author} {\bibinfo {author} {\bibfnamefont {A.}~\bibnamefont
  {Daddi-Moussa-Ider}}, \bibinfo {author} {\bibfnamefont {R.}~\bibnamefont
  {Golestanian}},\ and\ \bibinfo {author} {\bibfnamefont {A.}~\bibnamefont
  {Vilfan}},\ }\bibfield  {title} {\bibinfo {title} {Minimum entropy production
  by microswimmers with internal dissipation},\ }\href
  {https://doi.org/10.1038/s41467-023-41280-z} {\bibfield  {journal} {\bibinfo
  {journal} {Nat. Commun.}\ }\textbf {\bibinfo {volume} {14}},\ \bibinfo
  {pages} {6060} (\bibinfo {year} {2023})}\BibitemShut {NoStop}%
\bibitem [{\citenamefont {Bebon}\ \emph {et~al.}(2024)\citenamefont {Bebon},
  \citenamefont {Robinson},\ and\ \citenamefont {Speck}}]{speck}%
  \BibitemOpen
  \bibfield  {author} {\bibinfo {author} {\bibfnamefont {R.}~\bibnamefont
  {Bebon}}, \bibinfo {author} {\bibfnamefont {J.~F.}\ \bibnamefont
  {Robinson}},\ and\ \bibinfo {author} {\bibfnamefont {T.}~\bibnamefont
  {Speck}},\ }\href {https://doi.org/10.48550/ARXIV.2401.02252} {\bibinfo
  {title} {Thermodynamics of active matter: Tracking dissipation across
  scales}} (\bibinfo {year} {2024})\BibitemShut {NoStop}%
\bibitem [{\citenamefont {Alder}\ and\ \citenamefont
  {Wainwright}(1967)}]{Alder1967}%
  \BibitemOpen
  \bibfield  {author} {\bibinfo {author} {\bibfnamefont {B.~J.}\ \bibnamefont
  {Alder}}\ and\ \bibinfo {author} {\bibfnamefont {T.~E.}\ \bibnamefont
  {Wainwright}},\ }\bibfield  {title} {\bibinfo {title} {Velocity
  autocorrelations for hard spheres},\ }\href
  {https://doi.org/10.1103/PhysRevLett.18.988} {\bibfield  {journal} {\bibinfo
  {journal} {Phys. Rev. Lett.}\ }\textbf {\bibinfo {volume} {18}},\ \bibinfo
  {pages} {988} (\bibinfo {year} {1967})}\BibitemShut {NoStop}%
\bibitem [{\citenamefont {Zwanzig}\ and\ \citenamefont
  {Bixon}(1970)}]{Zwanzig1970}%
  \BibitemOpen
  \bibfield  {author} {\bibinfo {author} {\bibfnamefont {R.}~\bibnamefont
  {Zwanzig}}\ and\ \bibinfo {author} {\bibfnamefont {M.}~\bibnamefont
  {Bixon}},\ }\bibfield  {title} {\bibinfo {title} {Hydrodynamic theory of the
  velocity correlation function},\ }\href
  {https://doi.org/10.1103/PhysRevA.2.2005} {\bibfield  {journal} {\bibinfo
  {journal} {Phys. Rev. A}\ }\textbf {\bibinfo {volume} {2}},\ \bibinfo {pages}
  {2005} (\bibinfo {year} {1970})}\BibitemShut {NoStop}%
\bibitem [{\citenamefont {Langevin}(1908)}]{Langevin_1908}%
  \BibitemOpen
  \bibfield  {author} {\bibinfo {author} {\bibfnamefont {P.}~\bibnamefont
  {Langevin}},\ }\bibfield  {title} {\bibinfo {title} {Sur la théorie du
  mouvement brownien},\ }\href@noop {} {\bibfield  {journal} {\bibinfo
  {journal} {Comptes-rendus de l’Académie des Sciences}\ }\textbf {\bibinfo
  {volume} {146}},\ \bibinfo {pages} {530–532} (\bibinfo {year}
  {1908})}\BibitemShut {NoStop}%
\bibitem [{\citenamefont {Golestanian}(2024)}]{Ramin_10.1051epn2024305}%
  \BibitemOpen
  \bibfield  {author} {\bibinfo {author} {\bibfnamefont {R.}~\bibnamefont
  {Golestanian}},\ }\bibfield  {title} {\bibinfo {title} {Non-reciprocal
  active-matter: a tale of “loving hate, brawling love” across the
  scales},\ }\href {https://doi.org/10.1051/epn/2024305} {\bibfield  {journal}
  {\bibinfo  {journal} {Europhysics News}\ }\textbf {\bibinfo {volume} {55}},\
  \bibinfo {pages} {12} (\bibinfo {year} {2024})}\BibitemShut {NoStop}%
\bibitem [{\citenamefont {Dill}\ and\ \citenamefont
  {Brenner}(1983)}]{Dill1983}%
  \BibitemOpen
  \bibfield  {author} {\bibinfo {author} {\bibfnamefont {L.~H.}\ \bibnamefont
  {Dill}}\ and\ \bibinfo {author} {\bibfnamefont {H.}~\bibnamefont {Brenner}},\
  }\bibfield  {title} {\bibinfo {title} {A general theory of taylor dispersion
  phenomena. vi. langevin methods},\ }\href
  {https://doi.org/10.1016/0021-9797(83)90419-8} {\bibfield  {journal}
  {\bibinfo  {journal} {Journal of Colloid and Interface Science}\ }\textbf
  {\bibinfo {volume} {93}},\ \bibinfo {pages} {343–365} (\bibinfo {year}
  {1983})}\BibitemShut {NoStop}%
\bibitem [{\citenamefont {Frankel}\ and\ \citenamefont
  {Brenner}(1989)}]{Frankel1989}%
  \BibitemOpen
  \bibfield  {author} {\bibinfo {author} {\bibfnamefont {I.}~\bibnamefont
  {Frankel}}\ and\ \bibinfo {author} {\bibfnamefont {H.}~\bibnamefont
  {Brenner}},\ }\bibfield  {title} {\bibinfo {title} {On the foundations of
  generalized taylor dispersion theory},\ }\href
  {https://doi.org/10.1017/s0022112089001679} {\bibfield  {journal} {\bibinfo
  {journal} {Journal of Fluid Mechanics}\ }\textbf {\bibinfo {volume} {204}},\
  \bibinfo {pages} {97–119} (\bibinfo {year} {1989})}\BibitemShut {NoStop}%
\bibitem [{\citenamefont {Illien}\ \emph {et~al.}(2017)\citenamefont {Illien},
  \citenamefont {Adeleke-Larodo},\ and\ \citenamefont
  {Golestanian}}]{Illien2017}%
  \BibitemOpen
  \bibfield  {author} {\bibinfo {author} {\bibfnamefont {P.}~\bibnamefont
  {Illien}}, \bibinfo {author} {\bibfnamefont {T.}~\bibnamefont
  {Adeleke-Larodo}},\ and\ \bibinfo {author} {\bibfnamefont {R.}~\bibnamefont
  {Golestanian}},\ }\bibfield  {title} {\bibinfo {title} {Diffusion of an
  enzyme: The role of fluctuation-induced hydrodynamic coupling},\ }\href
  {https://doi.org/10.1209/0295-5075/119/40002} {\bibfield  {journal} {\bibinfo
   {journal} {EPL (Europhysics Letters)}\ }\textbf {\bibinfo {volume} {119}},\
  \bibinfo {pages} {40002} (\bibinfo {year} {2017})}\BibitemShut {NoStop}%
\bibitem [{\citenamefont {Adeleke-Larodo}\ \emph {et~al.}(2019)\citenamefont
  {Adeleke-Larodo}, \citenamefont {Illien},\ and\ \citenamefont
  {Golestanian}}]{AdelekeLarodo2019}%
  \BibitemOpen
  \bibfield  {author} {\bibinfo {author} {\bibfnamefont {T.}~\bibnamefont
  {Adeleke-Larodo}}, \bibinfo {author} {\bibfnamefont {P.}~\bibnamefont
  {Illien}},\ and\ \bibinfo {author} {\bibfnamefont {R.}~\bibnamefont
  {Golestanian}},\ }\bibfield  {title} {\bibinfo {title} {Fluctuation-induced
  hydrodynamic coupling in an asymmetric, anisotropic dumbbell},\ }\bibfield
  {journal} {\bibinfo  {journal} {The European Physical Journal E}\ }\textbf
  {\bibinfo {volume} {42}},\ \href {https://doi.org/10.1140/epje/i2019-11799-5}
  {10.1140/epje/i2019-11799-5} (\bibinfo {year} {2019})\BibitemShut {NoStop}%
\bibitem [{\citenamefont {Agudo-Canalejo}\ and\ \citenamefont
  {Golestanian}(2020)}]{AgudoCanalejo2020}%
  \BibitemOpen
  \bibfield  {author} {\bibinfo {author} {\bibfnamefont {J.}~\bibnamefont
  {Agudo-Canalejo}}\ and\ \bibinfo {author} {\bibfnamefont {R.}~\bibnamefont
  {Golestanian}},\ }\bibfield  {title} {\bibinfo {title} {Diffusion and steady
  state distributions of flexible chemotactic enzymes},\ }\href
  {https://doi.org/10.1140/epjst/e2020-900224-3} {\bibfield  {journal}
  {\bibinfo  {journal} {The European Physical Journal Special Topics}\ }\textbf
  {\bibinfo {volume} {229}},\ \bibinfo {pages} {2791–2806} (\bibinfo {year}
  {2020})}\BibitemShut {NoStop}%
\bibitem [{\citenamefont {Najafi}\ and\ \citenamefont
  {Golestanian}(2004)}]{Najafi2004}%
  \BibitemOpen
  \bibfield  {author} {\bibinfo {author} {\bibfnamefont {A.}~\bibnamefont
  {Najafi}}\ and\ \bibinfo {author} {\bibfnamefont {R.}~\bibnamefont
  {Golestanian}},\ }\bibfield  {title} {\bibinfo {title} {Simple swimmer at low
  reynolds number: Three linked spheres},\ }\href
  {https://doi.org/10.1103/PhysRevE.69.062901} {\bibfield  {journal} {\bibinfo
  {journal} {Phys. Rev. E}\ }\textbf {\bibinfo {volume} {69}},\ \bibinfo
  {pages} {062901} (\bibinfo {year} {2004})}\BibitemShut {NoStop}%
\bibitem [{\citenamefont {Golestanian}\ and\ \citenamefont
  {Ajdari}(2008{\natexlab{b}})}]{RGAA2008}%
  \BibitemOpen
  \bibfield  {author} {\bibinfo {author} {\bibfnamefont {R.}~\bibnamefont
  {Golestanian}}\ and\ \bibinfo {author} {\bibfnamefont {A.}~\bibnamefont
  {Ajdari}},\ }\bibfield  {title} {\bibinfo {title} {Analytic results for the
  three-sphere swimmer at low reynolds number},\ }\href
  {https://doi.org/10.1103/PhysRevE.77.036308} {\bibfield  {journal} {\bibinfo
  {journal} {Phys. Rev. E}\ }\textbf {\bibinfo {volume} {77}},\ \bibinfo
  {pages} {036308} (\bibinfo {year} {2008}{\natexlab{b}})}\BibitemShut
  {NoStop}%
\bibitem [{\citenamefont {Aditi~Simha}\ and\ \citenamefont
  {Ramaswamy}(2002)}]{Simha2002}%
  \BibitemOpen
  \bibfield  {author} {\bibinfo {author} {\bibfnamefont {R.}~\bibnamefont
  {Aditi~Simha}}\ and\ \bibinfo {author} {\bibfnamefont {S.}~\bibnamefont
  {Ramaswamy}},\ }\bibfield  {title} {\bibinfo {title} {Hydrodynamic
  fluctuations and instabilities in ordered suspensions of self-propelled
  particles},\ }\href {https://doi.org/10.1103/PhysRevLett.89.058101}
  {\bibfield  {journal} {\bibinfo  {journal} {Phys. Rev. Lett.}\ }\textbf
  {\bibinfo {volume} {89}},\ \bibinfo {pages} {058101} (\bibinfo {year}
  {2002})}\BibitemShut {NoStop}%
\bibitem [{\citenamefont {Saintillan}\ and\ \citenamefont
  {Shelley}(2008)}]{shelley2008}%
  \BibitemOpen
  \bibfield  {author} {\bibinfo {author} {\bibfnamefont {D.}~\bibnamefont
  {Saintillan}}\ and\ \bibinfo {author} {\bibfnamefont {M.~J.}\ \bibnamefont
  {Shelley}},\ }\bibfield  {title} {\bibinfo {title} {Instabilities and pattern
  formation in active particle suspensions: Kinetic theory and continuum
  simulations},\ }\href {https://doi.org/10.1103/PhysRevLett.100.178103}
  {\bibfield  {journal} {\bibinfo  {journal} {Phys. Rev. Lett.}\ }\textbf
  {\bibinfo {volume} {100}},\ \bibinfo {pages} {178103} (\bibinfo {year}
  {2008})}\BibitemShut {NoStop}%
\bibitem [{\citenamefont {Baskaran}\ and\ \citenamefont
  {Marchetti}(2009)}]{Baskaran2009}%
  \BibitemOpen
  \bibfield  {author} {\bibinfo {author} {\bibfnamefont {A.}~\bibnamefont
  {Baskaran}}\ and\ \bibinfo {author} {\bibfnamefont {M.~C.}\ \bibnamefont
  {Marchetti}},\ }\bibfield  {title} {\bibinfo {title} {Statistical mechanics
  and hydrodynamics of bacterial suspensions},\ }\href
  {https://doi.org/10.1073/pnas.0906586106} {\bibfield  {journal} {\bibinfo
  {journal} {Proceedings of the National Academy of Sciences}\ }\textbf
  {\bibinfo {volume} {106}},\ \bibinfo {pages} {15567–15572} (\bibinfo {year}
  {2009})}\BibitemShut {NoStop}%
\bibitem [{\citenamefont {Leoni}\ and\ \citenamefont
  {Liverpool}(2010)}]{Tannie2010}%
  \BibitemOpen
  \bibfield  {author} {\bibinfo {author} {\bibfnamefont {M.}~\bibnamefont
  {Leoni}}\ and\ \bibinfo {author} {\bibfnamefont {T.~B.}\ \bibnamefont
  {Liverpool}},\ }\bibfield  {title} {\bibinfo {title} {Swimmers in thin films:
  From swarming to hydrodynamic instabilities},\ }\href
  {https://doi.org/10.1103/PhysRevLett.105.238102} {\bibfield  {journal}
  {\bibinfo  {journal} {Phys. Rev. Lett.}\ }\textbf {\bibinfo {volume} {105}},\
  \bibinfo {pages} {238102} (\bibinfo {year} {2010})}\BibitemShut {NoStop}%
\bibitem [{\citenamefont {Brady}\ and\ \citenamefont
  {Bossis}(1988)}]{Brady1988}%
  \BibitemOpen
  \bibfield  {author} {\bibinfo {author} {\bibfnamefont {J.~F.}\ \bibnamefont
  {Brady}}\ and\ \bibinfo {author} {\bibfnamefont {G.}~\bibnamefont {Bossis}},\
  }\bibfield  {title} {\bibinfo {title} {Stokesian dynamics},\ }\href
  {https://doi.org/10.1146/annurev.fl.20.010188.000551} {\bibfield  {journal}
  {\bibinfo  {journal} {Annual Review of Fluid Mechanics}\ }\textbf {\bibinfo
  {volume} {20}},\ \bibinfo {pages} {111–157} (\bibinfo {year}
  {1988})}\BibitemShut {NoStop}%
\bibitem [{\citenamefont {Ishikawa}\ and\ \citenamefont
  {Pedley}(2008)}]{Pedley2008}%
  \BibitemOpen
  \bibfield  {author} {\bibinfo {author} {\bibfnamefont {T.}~\bibnamefont
  {Ishikawa}}\ and\ \bibinfo {author} {\bibfnamefont {T.~J.}\ \bibnamefont
  {Pedley}},\ }\bibfield  {title} {\bibinfo {title} {Coherent structures in
  monolayers of swimming particles},\ }\href
  {https://doi.org/10.1103/PhysRevLett.100.088103} {\bibfield  {journal}
  {\bibinfo  {journal} {Phys. Rev. Lett.}\ }\textbf {\bibinfo {volume} {100}},\
  \bibinfo {pages} {088103} (\bibinfo {year} {2008})}\BibitemShut {NoStop}%
\bibitem [{\citenamefont {Matas-Navarro}\ \emph {et~al.}(2014)\citenamefont
  {Matas-Navarro}, \citenamefont {Golestanian}, \citenamefont {Liverpool},\
  and\ \citenamefont {Fielding}}]{Fielding2014}%
  \BibitemOpen
  \bibfield  {author} {\bibinfo {author} {\bibfnamefont {R.}~\bibnamefont
  {Matas-Navarro}}, \bibinfo {author} {\bibfnamefont {R.}~\bibnamefont
  {Golestanian}}, \bibinfo {author} {\bibfnamefont {T.~B.}\ \bibnamefont
  {Liverpool}},\ and\ \bibinfo {author} {\bibfnamefont {S.~M.}\ \bibnamefont
  {Fielding}},\ }\bibfield  {title} {\bibinfo {title} {Hydrodynamic suppression
  of phase separation in active suspensions},\ }\href
  {https://doi.org/10.1103/PhysRevE.90.032304} {\bibfield  {journal} {\bibinfo
  {journal} {Phys. Rev. E}\ }\textbf {\bibinfo {volume} {90}},\ \bibinfo
  {pages} {032304} (\bibinfo {year} {2014})}\BibitemShut {NoStop}%
\bibitem [{\citenamefont {Hohenberg}\ and\ \citenamefont
  {Halperin}(1977)}]{halperin1977}%
  \BibitemOpen
  \bibfield  {author} {\bibinfo {author} {\bibfnamefont {P.~C.}\ \bibnamefont
  {Hohenberg}}\ and\ \bibinfo {author} {\bibfnamefont {B.~I.}\ \bibnamefont
  {Halperin}},\ }\bibfield  {title} {\bibinfo {title} {Theory of dynamic
  critical phenomena},\ }\href {https://doi.org/10.1103/RevModPhys.49.435}
  {\bibfield  {journal} {\bibinfo  {journal} {Rev. Mod. Phys.}\ }\textbf
  {\bibinfo {volume} {49}},\ \bibinfo {pages} {435} (\bibinfo {year}
  {1977})}\BibitemShut {NoStop}%
\bibitem [{\citenamefont {Tiribocchi}\ \emph {et~al.}(2015)\citenamefont
  {Tiribocchi}, \citenamefont {Wittkowski}, \citenamefont {Marenduzzo},\ and\
  \citenamefont {Cates}}]{activeH2015}%
  \BibitemOpen
  \bibfield  {author} {\bibinfo {author} {\bibfnamefont {A.}~\bibnamefont
  {Tiribocchi}}, \bibinfo {author} {\bibfnamefont {R.}~\bibnamefont
  {Wittkowski}}, \bibinfo {author} {\bibfnamefont {D.}~\bibnamefont
  {Marenduzzo}},\ and\ \bibinfo {author} {\bibfnamefont {M.~E.}\ \bibnamefont
  {Cates}},\ }\bibfield  {title} {\bibinfo {title} {Active model h: Scalar
  active matter in a momentum-conserving fluid},\ }\href
  {https://doi.org/10.1103/PhysRevLett.115.188302} {\bibfield  {journal}
  {\bibinfo  {journal} {Phys. Rev. Lett.}\ }\textbf {\bibinfo {volume} {115}},\
  \bibinfo {pages} {188302} (\bibinfo {year} {2015})}\BibitemShut {NoStop}%
\bibitem [{\citenamefont {Landau}\ and\ \citenamefont
  {Lifshitz}(2013)}]{landau2013fluid}%
  \BibitemOpen
  \bibfield  {author} {\bibinfo {author} {\bibfnamefont {L.~D.}\ \bibnamefont
  {Landau}}\ and\ \bibinfo {author} {\bibfnamefont {E.~M.}\ \bibnamefont
  {Lifshitz}},\ }\href@noop {} {\emph {\bibinfo {title} {Fluid Mechanics:
  Landau and Lifshitz: Course of Theoretical Physics}}},\ Vol.~\bibinfo
  {volume} {6}\ (\bibinfo  {publisher} {Elsevier},\ \bibinfo {year}
  {2013})\BibitemShut {NoStop}%
\bibitem [{\citenamefont {Kr\"uger}\ \emph {et~al.}(2024)\citenamefont
  {Kr\"uger}, \citenamefont {Asheichyk}, \citenamefont {Kardar},\ and\
  \citenamefont {Golestanian}}]{Kruger2024}%
  \BibitemOpen
  \bibfield  {author} {\bibinfo {author} {\bibfnamefont {M.}~\bibnamefont
  {Kr\"uger}}, \bibinfo {author} {\bibfnamefont {K.}~\bibnamefont {Asheichyk}},
  \bibinfo {author} {\bibfnamefont {M.}~\bibnamefont {Kardar}},\ and\ \bibinfo
  {author} {\bibfnamefont {R.}~\bibnamefont {Golestanian}},\ }\bibfield
  {title} {\bibinfo {title} {Scale-dependent heat transport in dissipative
  media via electromagnetic fluctuations},\ }\href
  {https://doi.org/10.1103/PhysRevLett.132.106903} {\bibfield  {journal}
  {\bibinfo  {journal} {Phys. Rev. Lett.}\ }\textbf {\bibinfo {volume} {132}},\
  \bibinfo {pages} {106903} (\bibinfo {year} {2024})}\BibitemShut {NoStop}%
\bibitem [{\citenamefont {Zinn-Justin}(2021)}]{ZinnJustin2021}%
  \BibitemOpen
  \bibfield  {author} {\bibinfo {author} {\bibfnamefont {J.}~\bibnamefont
  {Zinn-Justin}},\ }\href {https://doi.org/10.1093/oso/9780198834625.001.0001}
  {\emph {\bibinfo {title} {Quantum Field Theory and Critical Phenomena: Fifth
  Edition}}}\ (\bibinfo  {publisher} {Oxford University Press, Oxford},\
  \bibinfo {year} {2021})\BibitemShut {NoStop}%
\bibitem [{\citenamefont {Oseen}(1927)}]{oseen1927neuere}%
  \BibitemOpen
  \bibfield  {author} {\bibinfo {author} {\bibfnamefont {C.~W.}\ \bibnamefont
  {Oseen}},\ }\href@noop {} {\emph {\bibinfo {title} {Neuere Methoden und
  Ergebnisse in der Hydrodynamik}}},\ Vol.~\bibinfo {volume} {1}\ (\bibinfo
  {publisher} {Akademische Verlagsgesellschaft},\ \bibinfo {year}
  {1927})\BibitemShut {NoStop}%
\end{thebibliography}%

\end{document}